\documentstyle[preprint,aps,prb,epsf]{revtex}
 
\tightenlines  
\begin{document}
\draft
\title{Comparison of 1A GeV $^{\bf 197}$Au + C data
with  thermodynamics:
the nature of phase transition in nuclear multifragmentation}

\author{R. P. Scharenberg$^1$,
B. K. Srivastava$^1$,
S. Albergo$^2$,
F. Bieser$^6$,
F. P.  Brady$^3$,
Z. Caccia$^2$,\\
D. A. Cebra$^3$,
A. D. Chacon$^{7,(a)}$
J. L. Chance$^3$,
Y. Choi$^{1,(b)}$
S. Costa$^2$,\\
J. B. Elliott$^{1,(c)}$,
M. L. Gilkes$^{1,(d)}$,
J. A. Hauger$^{1,(e)}$
A. S. Hirsch$^1$,
E. L. Hjort$^1$,\\
A. Insolia$^2$,
M. Justice$^{5,(f)}$,
D. Keane$^5$,
J. C.  Kintner$^{3,(g)}$
V. Lindenstruth$^{4,(h)}$
M. A. Lisa$^{6,(i)}$
H. S. Matis$^6$,
M. McMahan$^6$,
C.  McParland$^{6,(d)}$,
W. F. J. M\"{u}ller$^4$,\\
D. L. Olson$^6$,
M. D. Partlan$^{3,(c)}$
N. T. Porile$^1$,
R.  Potenza$^2$,
G. Rai$^6$,\\
J. Rasmussen$^6$,
H. G. Ritter$^6$,
J. Romanski$^{2,(j)}$
J. L.  Romero$^3$,
G. V. Russo$^2$,\\
H. Sann$^4$,
A. Scott$^5$,
Y.  Shao$^{5,(k)}$
T. J. M. Symons$^6$,
M. Tincknell$^{1,(l)}$,\\
C. Tuv\'{e}$^2$,
S. Wang$^5$,
P. Warren$^{1,(m)}$
H. H. Wieman$^6$,
T. Wienold$^{6,(h)}$, and
K. Wolf$^{7,(n)}$\\
(EOS Collaboration)
}

\address{$^1$Purdue University, West Lafayette, IN 47907\\
$^2$Universit\'{a} di Catania and Istituto Nazionale di Fisica
Nucleare-Sezione di Catania,\\
95129 Catania, Italy\\
$^3$University of California, Davis, CA 95616\\
$^4$GSI, D-64220 Darmstadt, Germany\\
$^5$Kent State University, Kent, OH 44242\\
$^6$Nuclear Science Division, Lawrence Berkeley National Laboratory, 
Berkeley, CA 94720\\
$^7$Texas A\&M University, College Station, TX  77843}

\date{\today}

\maketitle

\newpage

\begin{abstract}
Multifragmentation (MF) results from 1A GeV Au on C have been compared
with the Copenhagen statistical multifragmentation model (SMM).  The
complete charge, mass and momentum reconstruction of the Au projectile
was used to identify high momentum ejectiles leaving an excited remnant
of mass $A$, charge $Z$ and excitation energy $E^{\ast}$ which
subsequently multifragments.  Measurement of the magnitude and
multiplicity (energy) dependence of the initial free volume and the
breakup volume determines the variable volume parameterization
of SMM.  Very good agreement is obtained using
 SMM with the standard values of the SMM  parameters. A large number of
observables, including the fragment charge yield distributions,
fragment multiplicity distributions, caloric curve, 
critical exponents, and the critical scaling function are
explored in this comparison.  The two stage structure of SMM is used to
determine the effect of cooling of the primary hot fragments.  Average
fragment yields with $Z\geq3$ are essentially unaffected when the
excitation energy is $\leq$ 7 MeV/nucleon.  SMM studies suggest that
the experimental critical exponents are largely unaffected by cooling
and event mixing.  The nature of the phase transition in SMM is studied
as a function of the remnant mass and charge using the microcanonical
equation of state.  For light remnants $A \leq $ 100, backbending
is observed
indicating  negative specific heat, while for  $A \geq$ 170 
the  effective latent
heat approaches zero. Thus for heavier systems this  
transition  can be identified as
 a continuous thermal phase transition  where a large nucleus breaks up
into a number of smaller nuclei with only a minimal release of
constituent nucleons. $Z \leq$ 2 particles are primarily emitted in the
initial collision and after MF in the fragment deexcitation process.
\end{abstract}


\begin{description}
\item
$^{(a)}$ 
Present address: Space Systems/Loral, Palo Alto, CA 94202.
\item
$^{(b)}$ 
Present address Sun Kwun Kwan University, Suwon
440-746, Republic of Korea.
\item
$^{(c)}$ 
Present address: Nuclear Science Division,
Lawrence Berkeley National Laboratory, Berkeley, CA 94720.
\item
$^{(d)}$ 
Present address: Mission Research Corp., Santa Barbara, CA 93102.
\item
$^{(e)}$ 
Present address: Augusta State College, Augusta, GA 30910.
\item
$^{(f)}$ 
Present address: Brookhaven National Laboratory, Upton, NY 11793.
\item
$^{(g)}$ 
Present address: St Mary's College, Morage, CA 94575.
\item
$^{(h)}$ 
Universit\"at Heidelberg, Germany.
\item
$^{(i)}$ 
Present address: Ohio State University, Columbus, OH 43210.
\item
$^{(j)}$
Present address: The Svedberg Laboratory,
University of Uppsala, S751-21, Uppsala, Sweden.
\item
$^{(k)}$
Present address: Crump Institute for Biological
Imaging, UCLA, Los Angeles, CA 91776.
\item
$^{(l)}$
MIT Lincoln Laboratory, Lexington, MA 02420.
\item
$^{(m)}$
Present address: Naval Research Laboratory, Washington, D.C. 20375.
\item
$^{(n)}$
Deceased.
\end{description} 

\newpage

\section{INTRODUCTION}

Multifragmentation (MF)  is the dominant decay mode in
heavy ion reactions when the excitation energy is
comparable to the nuclear binding energy.  The recent use of 4$\pi$
detectors capable of observing a substantial fraction of the particles and
fragments emitted in a given interaction has been invaluable in furthering
the elucidation of this complex phenomenon
\cite{hubele91,jeong94,deses93,steckmeyer96,morley96,peaslee94}. 
Several review articles of MF have been published in recent years
\cite{lynch87,moretto93,peilert94}.

One of the most complete MF experiments to date has been that performed
by the EOS collaboration for the $^{197}$Au on C system
\cite{gilkes94,hauger96,elliott96,hauger98,elliott98,lauret98,srivastava99,hauger00}.
The use of seamless detectors such as the EOS time projection chamber
coupled with the multiple sampling ionization chamber (MUSIC II)
permitted the observation of practically all the charged particles and
fragments emitted in each event, ranging from protons to heavy
fragments.  Full reconstruction was therefore possible for a large
fraction of the events.  The high energy asymmetric collision of a 1A
GeV projectile on a light target is uniquely favorable for the
kinematic separation of the initial nuclear collision and the
subsequent MF transition.  This data set permitted us to establish that
MF occurs following the expansion of a remnant formed with charge $Z$,
mass $A$ and excitation energy $E^{\ast}$ after the prompt ejection of
high momentum light particles \cite{hauger96,hauger98}.  We have
applied model independent methods used in the study of critical
phenomena to extract the value of several critical exponents
\cite{gilkes94,elliott96}.  The critical scaling function, which
describes the behavior of the nuclear remnant near a critical point,
was determined \cite{elliott98}.  In performing this analysis we were
aided by percolation calculations \cite{elliott94,elliott97}, which
served as a quantitative guide to the application of the methods
developed for the study of critical exponents in small systems.

The EOS experiment was an outgrowth of earlier inclusive studies of MF
in the interaction of Xe and Kr with high-energy protons
\cite{gaidos79,finn82,minich82,hirsch84,sangster87,mahi88,porile89}.
In this work high precision counter techniques were used to obtain
accurate information about the kinetic energy spectra of {\it
isotopically} resolved nuclear fragments.  The reduced Coulomb barrier
seen in the fragment spectra indicated that nuclear fragments are
emitted from an expanded nuclear system.  Systematics of the fragment
kinetic energy spectra also showed that the remnant was lighter than
the target nucleus implying that {\it the nuclear remnant existed
for a time} after the initial collision.  The relative fragment yields
of 63 isotopes could be understood using a thermal droplet model
\cite{fisher67} with a free energy parameterization based on the
semi-empirical mass formula.  For energy depositions of $\sim$8 MeV per
nucleon this multi-isotope thermometer gave a freezeout temperature of
$\sim$5 MeV.  The systematic variation of the ``kinetic temperature''
(exponential slope parameter extracted from fragment kinetic energy spectra) as a function of the
fragment mass also indicated that Fermi momentum in the remnant
must play a significant role in the fragment formation process.

More speculative was the suggestion that the power law yield of the fragment
masses Y=$A^{-\tau}$ with $\tau\sim$2.6 observed in 50-400 GeV p on Xe
collisions showed that nuclear fragmentation might give information about a
possible ``liquid-gas'' thermal phase transition in charged nuclear matter
\cite{finn82}.  Important support for this view came from the exclusive
emulsion data of Waddington and Freier \cite{wadd85}, which were analyzed
by Campi to show that the conditional moments of the individual fragment events exhibited
features characteristic of a critical transition \cite{campi86,campi88}.  The
occurrence of thermal equilibrium in the MF process was
strongly supported by the experiments of the ALADIN collaboration, which
showed that fragment yields were independent of the entrance channel when
the data was scaled for projectile or target  mass \cite{hubele91,schuttauf96}.

The above cited evidence for the formation of a substantially
equilibrated remnant which expands prior to MF suggests that it would
be appropriate to compare the EOS results with a thermal model in which
MF occurs from an expanded state.  Two widely used thermal
multifragmentation models are the Copenhagen (SMM) \cite{bondorf95} and
Berlin (MMMC) \cite{gross90} statistical treatments.  These models
differ in their parameterization of the expansion and in technical
details but are similar in their thermodynamic approach.  For very
small systems both these microcanonical models predict the onset of a
very inhomogeneous state at a definite excitation energy. This
inhomogeneous excited state consists of a number of normal density
fragments accompanied by a statistically insignificant number of
nucleons.  Subsequently these fragments cool by light particle
emission.  We present here a comparison of the EOS data with SMM.
Several data and model  comparisons have been made previously
\cite{hubele92,barz93,kreutz93,botvina95,des96,dagos96,xi97,williams97,campi94}.
These comparisons seriously suffered from the fact that the experiments
did not determine the $Z$, $A$, and   $E^*$ of the remnant.  Instead,
these $Z$, $A$,   $E^{\ast}$ values, which constitute the proper input
to test models of MF, were obtained either by use of a dynamical first
stage model \cite{hubele92,kreutz93}, by using a comparison of some of
the data with SMM to constrain them
\cite{barz93,botvina95,dagos96,xi97,williams97}, or by more complex
backtracing from fragment data \cite{des96,campi94}.  A great advantage
afforded by asymmetric reverse kinematics collisions is that they
permit an accurate separation of the initial reaction phase from the
subsequent decay of the excited remnant.

The use of SMM also permits us to examine several aspects of
the EOS critical exponent  analysis that are potentially problematic.  The EOS results
were obtained for fragments in their final, cold state.  However, 
the fragments are presumably formed in an expanded  hot state. 
In SMM the remnant is equilibrated, the fragments are formed in the hot
system,
then separate under
the influence of the Coulomb force and undergo deexcitation.  As has
been noted elsewhere \cite{williams97}, the distribution of the hot,
primary fragments may be affected by secondary decay.  The difference
between the two distributions could, in principle, affect the values of
the critical exponents.  Since SMM gives separate results for hot and
cold fragments, the effects of secondary decay can be probed.

The EOS results indicate that the first prompt step leads to a distribution
of remnants.  The analysis groups these remnants
according to the total charged particle multiplicity, $m$, which serves as
the control parameter.  Events characterized by a given multiplicity
will generally include a range of remnant $Z, A,  E^{\ast}$ 
values.  It has
been noted that such event mixing can affect the values of the critical
exponents \cite{bauer95}.  Furthermore, the use of multiplicity instead
of temperature or excitation energy as the control parameter has been
questioned \cite{bauer2,mastinu98}.  Comparison of SMM results
obtained for the experimental distribution of remnants with those obtained
for a single remnant permits us to probe the importance of these effects.

This paper is organized in the following manner.  Section II gives a
brief summary of SMM and  shows how the data determine the variable volume
of SMM.  Only a single adjustable parameter remains.
Section III summarizes
the properties of the experimental remnant distribution, which serves
as the input data to SMM.  The occurrence of radial expansion energy as
part of the excitation energy is discussed in this section.  The
comparison between experimental and model results is presented in
Section IV.  The various factors that can affect the fragment yields,
the extraction of critical exponents, and the critical multiplicity 
 are considered in Section V.
In Sec. VI,  the physics of MF and the
nature of the thermal phase transition in SMM as a function
of the remnant mass and charge is explored. 
The identification of the order of a thermal phase transition
for the  MF of Au on C requires the use  of the
microcanonical equation of state. 
A summary of the results and our
conclusions are given in Section VII.

\section{THE SMM MODEL}

SMM is a statistical description of the simultaneous breakup of an
expanded excited nucleus into nucleons and hot fragments \cite{bondorf95}. 
Individual fragments at normal nuclear density are described with a charged
liquid drop parameterization.  The free energy of a fragment 
$A, \; Z \; (Z \geq 3)$
is given by
\begin{equation}
F_{A ,Z}~=~F_{trans} + F_{vol}+F_{surf} +F_{sym} +
F_{Coul}
\end{equation}
and is used to determine the fragment formation probability.  
This solution explicitly  assumes the inhomogeneous nature of the hot MF final state.
Light fragments $Z < 3$ may also be present in the hot MF final state.
For the $Z \geq 3$ fragments, a
quantum mechanical description is used  for the temperature dependent
 volume, surface,  and translational  free energy of the fragments.
The temperature independent  parameters are based on the coefficients of the semiempirical mass formula.
The critical temperature, at which the surface tension of 
neutral nuclear matter droplets would go to zero,
is in the range suggested by infinite neutral nuclear matter calculations
\cite{ravenhall83}.  

In SMM the  translational free energy 
depends on the {\it free} volume. 
The free volume, $V_{f}$,  can be expressed in terms of
the volume of the multifragmenting system at normal nuclear density, $V_{rem}$, 
\begin{equation}
V_f ~=~\chi V_{rem}
\end{equation}
where the free volume parameter $\chi$ depends on the SMM fragment
multiplicity  according to the relation: 
\begin{equation}
\chi~=~\left[ 1 + ~{d \over R_o}~(M^{\frac{1}{3}}-1)\right]^3 -1
\end{equation}
where $R_o$=1.17 $A_o^{1/3}$ fm  
and $M$ is the {\it charged plus neutral} hot fragment multiplicity.
The crack width parameter,
$d$, scales the magnitude of the  multiplicity
dependent free volume.  
The breakup volume $V_{b}$, which includes the
volume of the fragments, is $V_{b} = (1+ \kappa )V_{rem}$, where $\kappa$ is the
Coulomb reduction parameter\cite{bondorf95}.  

We have previously shown that energy deposition in the Au on C reaction
is proportional to nucleons knocked out of the Au nucleus by quasi-elastic
nucleon-nucleon scattering \cite{hauger98}.  If we assume that the excited
remnant initially is produced in the Au volume, then the experimentally
determined {\it initial free volume} is given by:
\begin{equation}
V_{f}^{i} = V_{Au} \; 
(A_{Au} - A_{rem} )/
A_{Au}  
\end{equation}
and the remnant volume is given by:
\begin{equation}
V_{rem} = V_{Au}\;  (A_{rem}/A_{Au})  
\end{equation}
Proton-proton correlation experiments
for the multifragmentation of 1A GeV Au on Au show that the freeze out volume
is $\sim$2 V$_{Au}$,  nearly independent of excitation energy  and remnant mass and charge \cite{fritz97}.
\begin{figure}
\epsfxsize=8.5cm
\centerline{\epsfbox{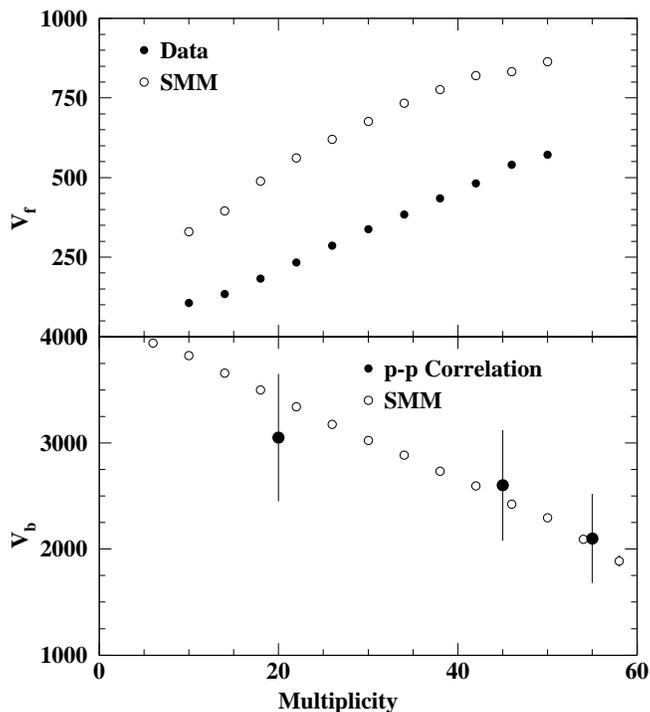}}
\vspace*{0.003in}
\caption{a) SMM free volume (Eq. 2) and the experimental initial free
volume (Eq. 4) for the MF region. 
b) SMM breakup volume and the  experimental freeze out volume from $p$-$p$
correlations. The variation in $V_{b}$ is due to the decrease of the remnant
mass. The remnant does not expand in compound nucleus events which
occur at low multiplicity.
}
\label{fig.1}
\end{figure}
Fig.1a shows the SMM free volume $V_{f}$, 
 and  the initial free volume $V_{f}^{i}$
as a function of the multiplicity.
As expected, the SMM free volume  $V_{f}$ is about twice the initial free
volume, consistent with the  expansion from $V_{Au}$ to $2V_{Au}$.
Note that the slopes of the $V_{f}$ and $V_{f}^{i}$ versus multiplicity curves
track over the MF region. 

Fig.~1b compares the breakup volume 
\begin{equation}
V_{b} = (1 + \kappa) V_{rem} = (1 + \kappa) 
(A_{rem} /A_{Au})\; V_{Au}
\end{equation}
with the freeze out volume from $p$-$p$ correlations 
as a function of the multiplicity. 
The breakup volume $V_{b}$ tracks the
freezeout volume from the $p$-$p$ correlation experiments.
The freeze out volume calculated in our earlier publication \cite{hauger98} is
different 
as it was obtained using initial volume of Au nucleus and not the remnant 
volume.  
Thus, experiment confirms the
structure (M dependence)  and scale (crack width parameter $d$)
 of the volume parameterization of SMM for the 1A GeV Au
on C experiment.
In addition to these experimental arguments, theoretical arguments based
on BUU modeling of the nuclear collision \cite{daniel95} also suggest that the
expansion of the remnant is energy dependent.

The remaining parameter is $\epsilon_{0}$, the inverse level density
parameter.  The $\epsilon_{0}$ 
values were determined by comparison of SMM with the various
experimental fragment yield distributions  and it was found that
$\epsilon_{0} = 16$ MeV.
 Thus, the so-called standard values of  all these parameters turn out to give the
best agreement with the data. (See Table~I.)

In SMM the primary fragments are propagated in their mutual
Coulomb field and then undergo secondary decay.
In the model successive particle emission 
from hot fragments with A$>$16 is assumed the  deexcitation mechanism.
 The deexcitation of these
fragments is treated by means of the standard Weisskopf evaporation model. 
Light fragments (A $<$ 16) deexcite via
Fermi breakup.  The lightest particles (A$<$4) can be formed only in their
ground states and undergo no secondary decay.  
We have used a version
of the model that incorporates only thermal degrees of freedom. 
Consequently, radial expansion or angular momentum are not included in this
comparison between data and  SMM.

\section{THE INPUT DATA}

\subsection{Properties of the remnant}

The reverse kinematic  EOS experiment permitted the identification of
charged projectile fragments $\rm 1 \leq Z \leq 79$ in 1A GeV \ 
${}^{197}$Au on C interactions with high efficiency.  The momenta of these
fragments were measured and used to decompose the reaction into a
prompt 
stage, in which high momentum Z=1, Z=2  fragments and neutrons are
emitted, and a second stage involving the decay of the remnant left
after this stage \cite{hauger98}.  The analysis presented
here is based on about 32,000 fully reconstructed MF events for which the total
charge of the reconstructed Au system was found to be 79$\pm$4.
Average fragment mass values  for a given Z were 
determined and used to reconstruct the MF final state mass $A'$
of the charged fragments. The number of free neutrons in the MF final state 
is used in the determination of $E^{\ast}$ by energy balance.

The remnant resulting from the prompt stage can be
characterized by Z, A, and $E^*$.  We follow previous practice and
report $E^*$ on a per nucleon basis.  The determination of these
quantities has been described in detail elsewhere \cite{hauger98}. Fig.2
shows the mass distribution of the remnants.
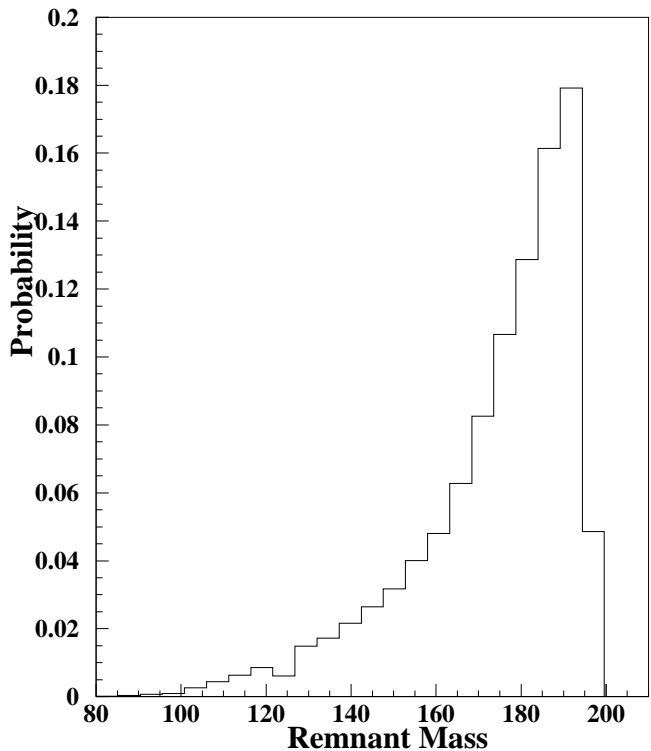
\begin{figure}
\epsfxsize=8.5cm
\centerline{\epsfbox{fig2.epsi}}
\vspace*{0.003in}
\caption{Mass distribution of the experimental remnant.
}
\label{fig.2}
\end{figure}
The most probable
mass is A$\sim$190.  However, the distribution is broad and extends
down to $A \sim 100$.  The $E^*$ distribution is shown in Fig.3a.
The distribution peaks at $\sim$2 MeV/nucleon but extends beyond 16
MeV/nucleon. 

It should be noted that the data were obtained with a
minimum bias trigger that eliminated some events with very low $E^*$.
These events do not lead to MF.  Fig. 4 shows the variation of
$E^*$ and A with  the total charged particle
 multiplicity $m$.  These quantities vary in the opposite
way with $m$, with $E^*$ increasing with $m$, and A showing a
concomitant decrease.  At a given $m$ there is a distribution in the
values of A and $E^*$.  For a given $m$, the width of the
distribution in A increases  from $\sim 1\%$ to $\sim 13\%$ while that
of the $E^*$ distribution is $\sim 25\%$ over most of the multiplicity
range \cite{hauger98}.
\begin{figure}
\epsfxsize=8.5cm
\centerline{\epsfbox{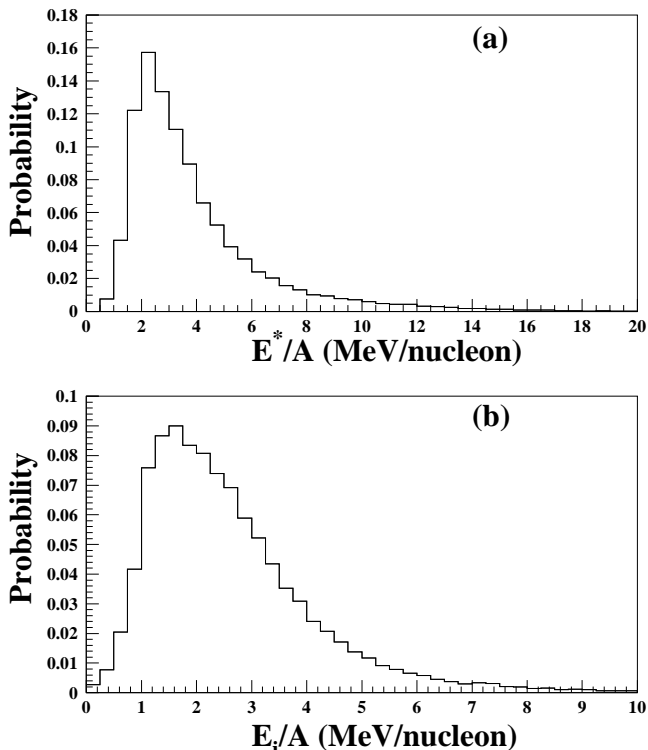}}
\vspace*{0.003in}
\caption{Distribution of remnants as a function of  (a) total excitation energy and (b) input excitation energy to SMM.
}
\label{fig.3}
\end{figure}
\subsection{The effect of non-thermal expansion}

We have previously shown that some of the excitation energy of the
remnant actually consists of nonthermal expansion energy, $E_x$
\cite{hauger98,lauret98}.  
The model input  energy $E_{i}$ is obtained by subtracting $E_{x}$  
from the excitation energy $E^{\ast}$.
Standard SMM does not include radial expansion and so only
$E_i$ must be used in the input data.  Fig.4 shows the
dependence of $E_x$ and $E_i$ on the multiplicity $m$. The expansion
energy is very small for $m{< \atop \sim} 20$ but then increases
sharply, becoming comparable to the input energy for the
largest observed multiplicity, $m\sim 60$.  The spectrum of $E_i$
values is plotted in Fig. 3b.  The distribution peaks at slightly less
than 2 MeV/nucleon and extends to $\sim$10 MeV/nucleon.
\begin{figure}
\epsfxsize=8.5cm
\centerline{\epsfbox{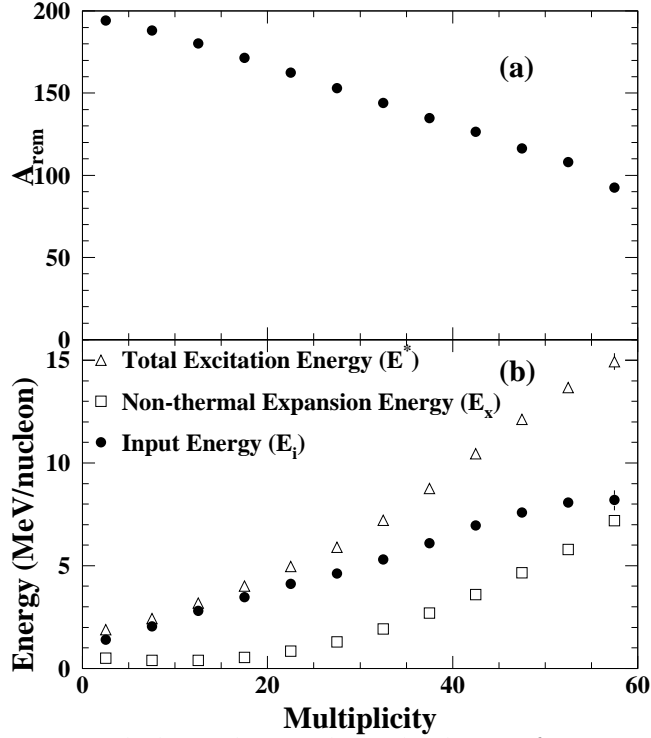}}
\vspace*{0.003in}
\caption{Dependence on total charged particle multiplicity of some average
properties of the experimental remnants:  (a) mass, (b) total excitation
energy ($E^*$), nonthermal expansion energy $(E_x)$, and input energy $(E_i )$.
All energies are in MeV/nucleon.
}
\label{fig.4}
\end{figure}
The expansion energy was obtained as the difference between the sum of
the measured charged  fragment mean kinetic energies
 and the translational thermal and
Coulomb energies of the fragments \cite{hauger98,warren96}.  This procedure
involves the use of the Albergo double isotope
ratio thermometer \cite{albergo85} 
to determine the {\it  isotopic
temperature} $T_a$.  This temperature determines the translational
thermal energy per particle, which is 3/2 $T_a$.  As shown in the next
section, SMM then  independently predicts the observed double isotope ratio.
This self consistency supports the validity of the $E_i$ determination
and permits a combined definitive test of the isotope ratio
thermometer $T_{a}$ and SMM.

Additional independent evidence for the presence of expansion energy in the
data can be seen in a comparison of the mean transverse kinetic
energies of fragments with the SMM predictions\cite{warren96}. 
Fig. 5 shows the
results for Li-N fragments for five multiplicity bins: 
1-9, 10-19, 20-29, 30-39, and 40-59.  Generally good agreement is obtained for
low multiplicities, as expected, because of the small contribution of the radial
expansion energy for small $m$ (see Fig.4).
The SMM transverse kinetic energy values decrease with increasing $m$.
\begin{figure}
\epsfxsize=8.5cm
\centerline{\epsfbox{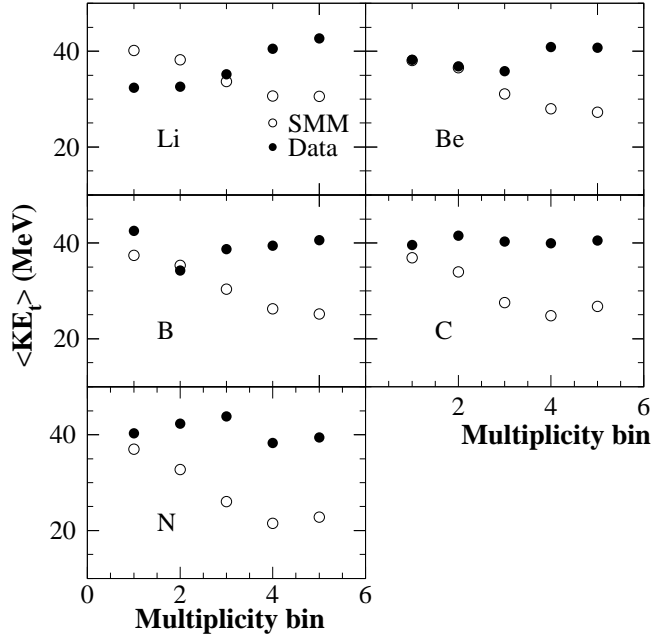}}
\vspace*{0.003in}
\caption{Mean transverse kinetic energies of fragments as a function of
nuclear charge. The  multiplicity bins are: 1-9, 10-19, 20-29, 30-39, and 
40-59, respectively.
}
\label{fig.5}
\end{figure}

In general, the SMM
transverse energies are smaller than the experimental
values.  The trend in the SMM values is the result of two factors, both
of which lead to a decrease in the translational kinetic energy with
increasing multiplicity:  (1) the increase in the volume occupied by
the fragments \cite{bondorf95} and (2) the decrease in the average
charge of the fragments.  In particular, see Fig.11, which
shows the variation of the calculated and experimental charges of the
largest fragment with $m$.  SMM is in excellent agreement with
experiment indicating that the reason for the discrepancy shown in
Fig.5 does not lie in the 
determination of the  mutual
Coulomb energy. 
Rather, the
increasing contribution of non-thermal expansion to the experimental energies
dominates the change  in Coulomb energy.  A similar
discrepancy has been observed when another data set was used to compare
SMM with
the Berlin statistical model \cite{gross90}.  Here too, the discrepancy
was attributed to radial expansion energy \cite{lauret98}. 
\section{COMPARISON OF EXPERIMENTAL DATA WITH SMM}

A single SMM calculation was performed for each of the 32,000 EOS
events.  The input data consisted of the Z, A, and $E_i$ values of the
remnants.  The output of each SMM calculation gives the Z and A values
of each fragment in its asymptotic cold state.  The distribution of these
MF products is used to make the comparisons presented in this section.
\begin{figure}
\epsfxsize=8.5cm
\centerline{\epsfbox{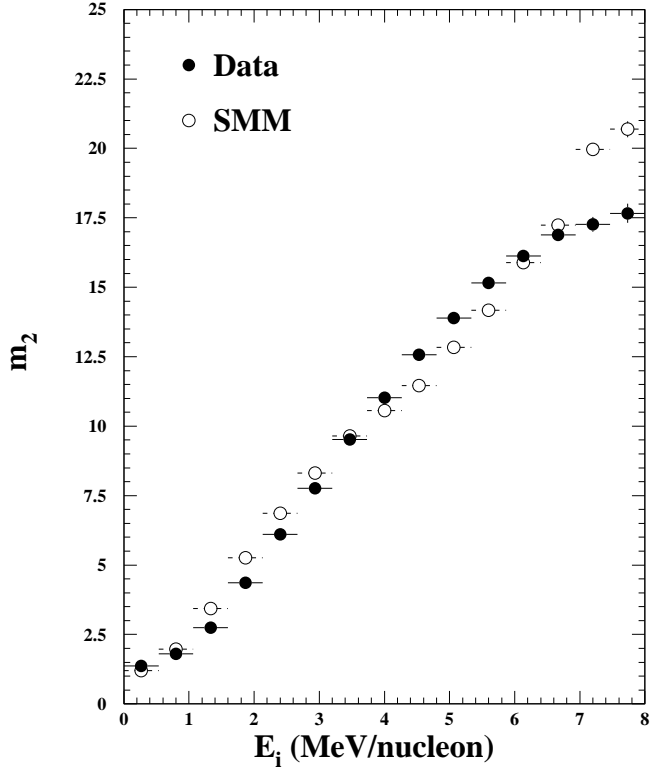}}
\vspace*{0.003in}
\caption{Variation of second stage multiplicity with input energy.
}
\label{fig.6}
\end{figure}

\subsection{Fragment yield and multiplicity comparisons}

Owing to the important role of multiplicity in the EOS experiment, we
compare the SMM and experimental multiplicities at the outset.  This
comparison is made as a function of $E_i$. See Fig. 6.  Since SMM does
not include a prompt first stage, the SMM charged particle multiplicity is
compared to the experimental second stage multiplicity, $m_2$.  This
quantity is obtained from $m$ by subtracting event-by-event the prompt first
stage multiplicity, $m_1$ \cite{hauger98}.  The calculated and experimental
$m_{2}$
distributions are in close agreement.  Because of this agreement, SMM results
will be plotted as a function of $m$ in many subsequent comparisons with
data.  Here $m$ will be the sum of SMM $m_2$ and experimental $m_1$
values.
\begin{figure}
\epsfxsize=8.5cm
\centerline{\epsfbox{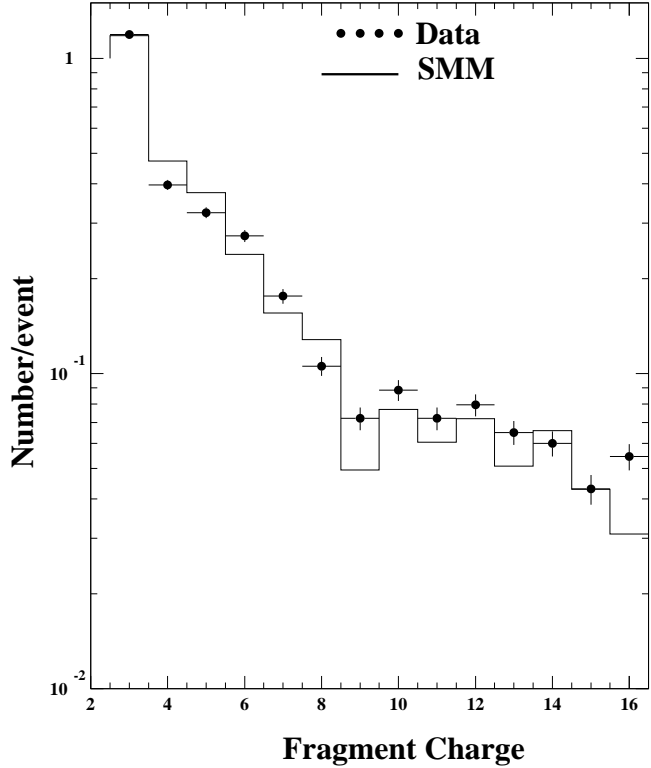}}
\vspace*{0.003in}
\caption{Yield of fragments averaged over all  $m$ 
as a function of fragment charge.
}
\label{fig.7}
\end{figure}
The fragment yield distributions offer the most direct comparison between
the data and SMM.  Fig.7 shows the number of Z=3-16
fragments per event averaged over all multiplicities.  Good agreement is
obtained for nearly all the fragments and the overall trend of decreasing
yield with increasing charge is well reproduced.  Both data and SMM show that
the yield of fluorine is suppressed relative to that of neighboring
fragments.  SMM indicates that this low yield reflects the influence of
final state interactions on the primary fragments.

The individual fragment yields are plotted as a function of $m$ in Figs.
8 and 9.  SMM generally does an excellent job of
reproducing the data except for the lightest fragments, where the model
predicts too many fragments for large $m$.  Both data and SMM show the
characteristic rise and fall of the individual
 fragment yields as a function of $m$ or
$E_i$.  Both of these distributions were used in the statistical analyses which extract
the power law and critical scaling behavior.  The 
average value of the 
total number of intermediate mass fragments $<IMF>$
as a function of $m$
is shown in Fig. 10. 
\begin{figure}
\epsfxsize=8.5cm
\centerline{\epsfbox{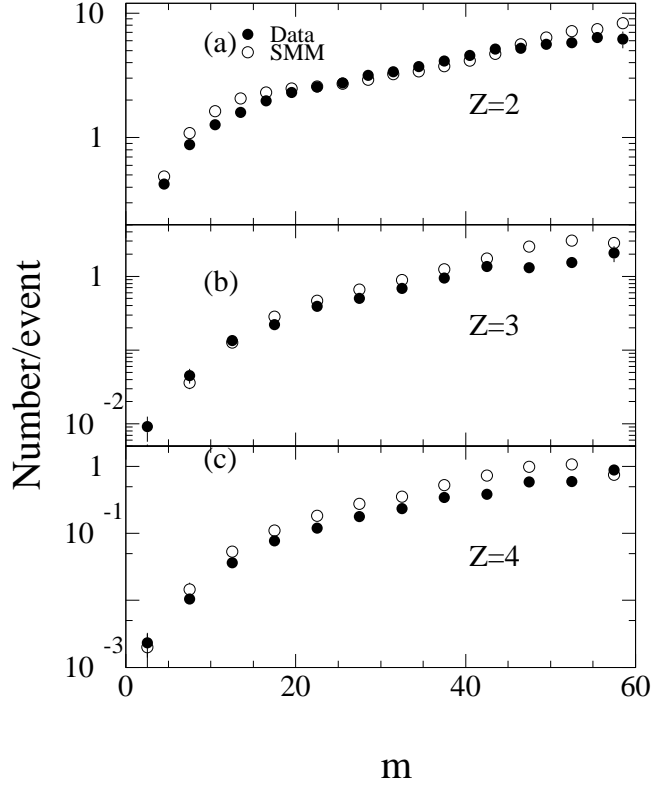}}
\vspace*{0.003in}
\caption{Average yield of fragments, for Z = 2, 3, 4 as a function of $m$.
}
\label{fig.8}
\end{figure}
The yield of IMF's
increases to $\sim$4.4/event at $m\sim 48$ and decreases for larger $m$. 
This behavior is reproduced by SMM, although the number of IMF's at
the peak is somewhat overestimated.  This difference is a consequence of
the above discrepancy between data and SMM for Li and Be fragments at
large $m$.  
Fig. 11 shows the average value of the charge of the largest fragment
in the distribution, $Z_{max}$.  SMM  gives the best agreement with the
data   using $\epsilon_{0} = 16$ MeV.
\begin{figure}
\epsfxsize=8.5cm
\centerline{\epsfbox{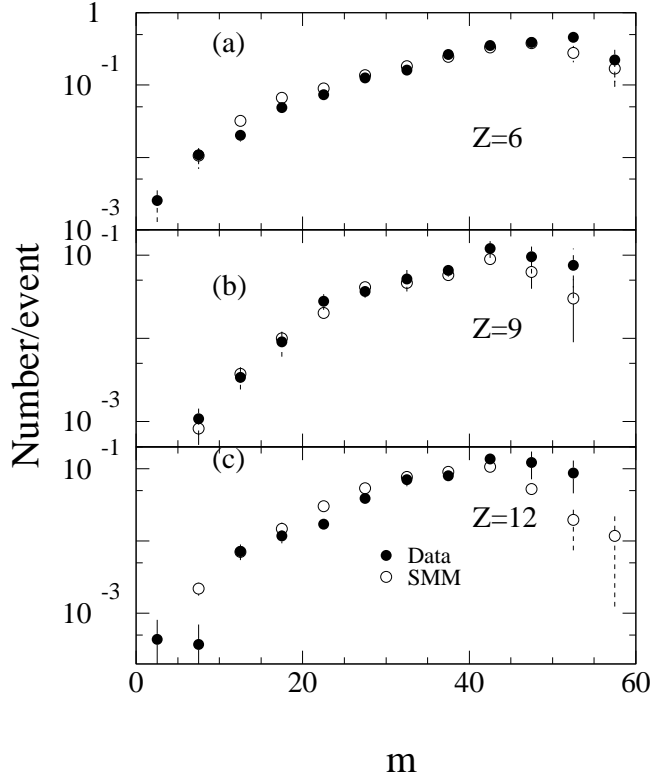}}
\vspace*{0.003in}
\caption{Average yield of fragments, for  Z=6, 9, 12 as a function
of $m$.
}
\label{fig.9}
\end{figure}
We have previously \cite{hauger98} obtained from the data the
isotopic temperature $T_{a}$  \cite{albergo85} on the basis of $^2$H/$^3$H and
$^3$He/$^4$He double isotopic yield ratios.  A value of $T_a$ was
also obtained from the $^6$Li/$^7$Li and $^3$He/$^4$He yield ratios.
Although the two values of $T_a$ are nearly equal, the $T_{HeDT}$
values were found to be more robust \cite{hauger98}.  These
temperatures are compared with the SMM values in Fig. 12.
Excellent agreement in the MF region is observed further confirming the self-consistent
nature of data and the predictions of SMM.

 It should be noted that the determination of freeze-out temperatures
obtained by the double isotope yield ratios is subject to correction due to
formation of these isotopes in secondary decay. This subject has been 
investigated by a number of workers \cite{xi97,cal1,cal2,cal3,cal4,cal5,cal6}.
For the  $T_{HeDT}$ thermometer the correction has been reported
$\sim$ 10\% below $E_{i}$ $\sim$  7 MeV/nucleon \cite{hauger98}. 

%
\begin{figure}
\epsfxsize=8.5cm
\centerline{\epsfbox{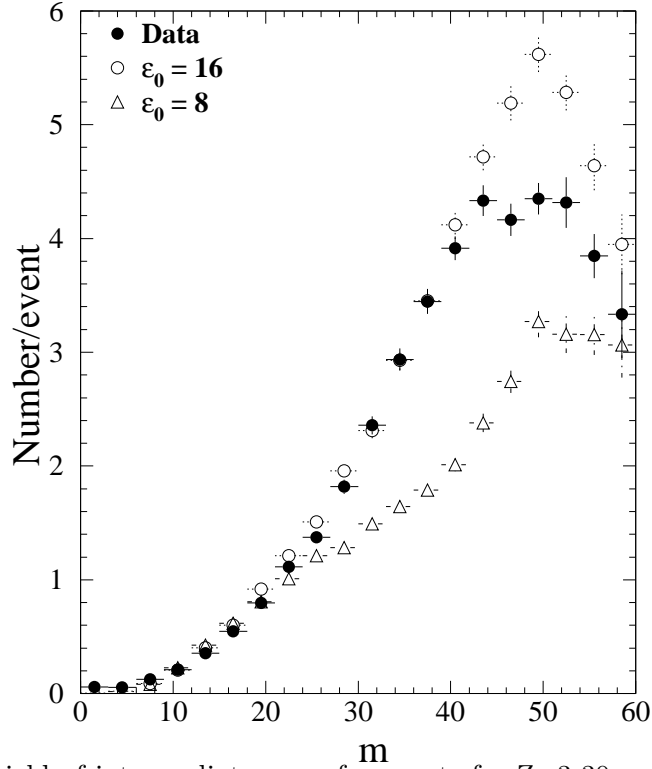}}
\vspace*{0.003in}
\caption{Average yield of intermediate mass fragments for  Z=3-30 as a 
 function of $m$.  SMM results are shown for $\epsilon_o$ = 16 and $\epsilon_o$ = 8.
}
\label{fig.10}
\end{figure}

\begin{figure}
\epsfxsize=8.5cm
\centerline{\epsfbox{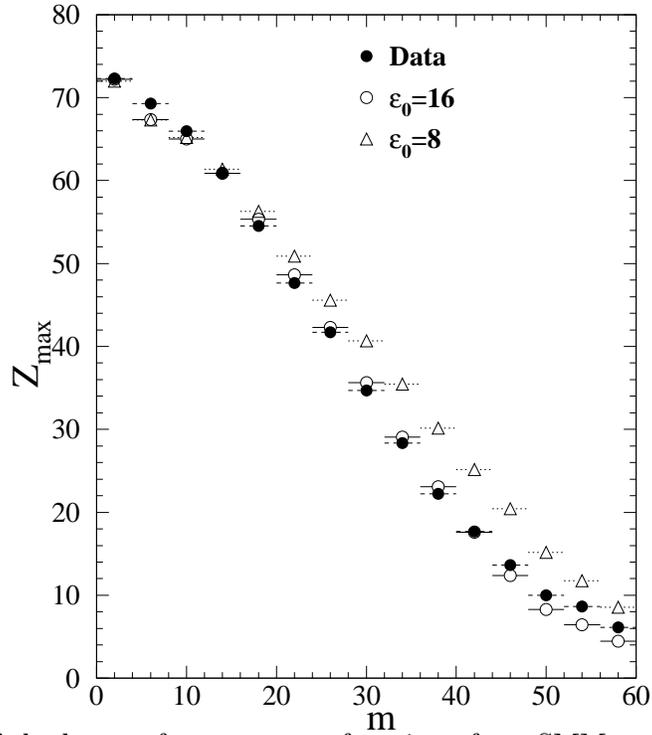}}
\vspace*{0.003in}
\caption{Charge of the largest fragment as a function of $m$.
SMM results are shown for $\epsilon_o$ = 16 and $\epsilon_o$ = 8.
}
\label{fig.11}
\end{figure}

\subsection{The caloric curve using isotopic temperatures}

The asymptotic caloric curve, 
which is a plot of  fragment
isotopic temperatures versus input energy, was first  obtained by
Pochodzalla et al. for 600 A MeV Au + Au collisions \cite{poch95}.  It
was found that the temperature had the essentially constant
value of $\sim$5 MeV for excitation energies between 3 and 10 MeV per
nucleon, a result that was interpreted as evidence for a first-order
phase transition.  A similar analysis of the EOS data, in which
the excitation energy included the expansion energy, showed that the
temperature increased continuously but slowly with $E^*$ over the above
range, i.e. from $\sim$4 to $\sim$6 MeV \cite{hauger98}.  Having
established the presence of non-thermal expansion, we can redetermine our
caloric curve and compare it with SMM.
\begin{figure}
\epsfxsize=8.5cm
\centerline{\epsfbox{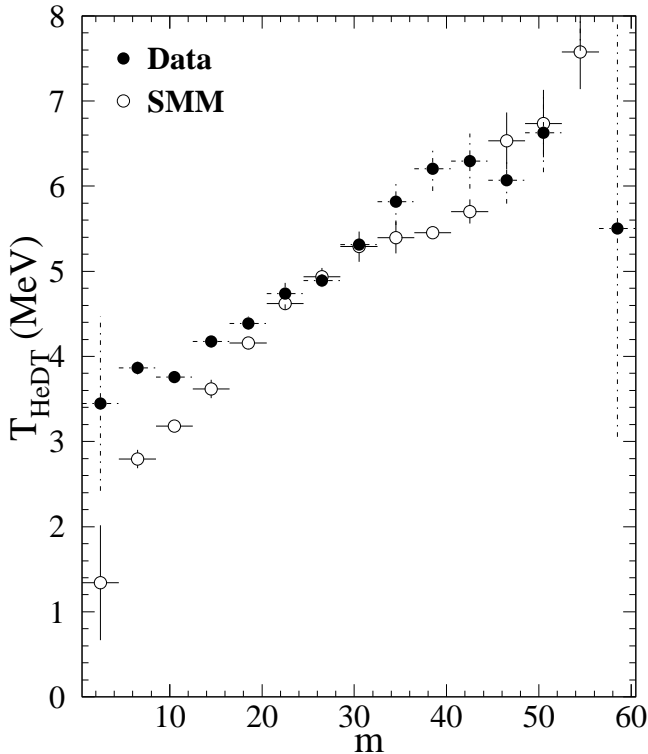}}
\vspace*{0.003in}
\caption{Dependence of the  $\rm {}^{2}H/^{3}H$ to $\rm {}^{4}He/^{3}He$
$\rm (T_{HeDT})$
isotopic ratio temperature on $m$.
}
\label{fig.12}
\end{figure}

These SMM and EOS caloric curves are compared in Fig. 13, where both SMM  isotopic temperatures and experimental 
isotopic temperatures are plotted
versus the experimental input energy $E_{i}$ per nucleon.  The SMM and EOS curves are
in close agreement and show a  very sharp increase in  the temperature T over
the experimental energy range.  Included for comparison is the
caloric curve previously obtained with the inclusion of
the expansion energy, which shows a much slower variation of T with $E^*$.  The
determination of a caloric curve in the manner proposed in
Ref. \cite{poch95} has already been shown to be problematic
\cite{natowitz95,campi96,moretto96,serfling98}.
 The recent reanalysis by the ALADIN group is in close agreement with our 
data\cite{muller99}.

  It must be noted that experimental caloric curves are actually inadequate
measures of the thermodynamic caloric curve, which involves the breakup 
temperature rather than the isotope ratio temperature. This has been discussed 
both by Bondorf et al. \cite{bondorf98} and in a recent publication from 
our group \cite{hauger00}.
\begin{figure}
\epsfxsize=8.5cm
\centerline{\epsfbox{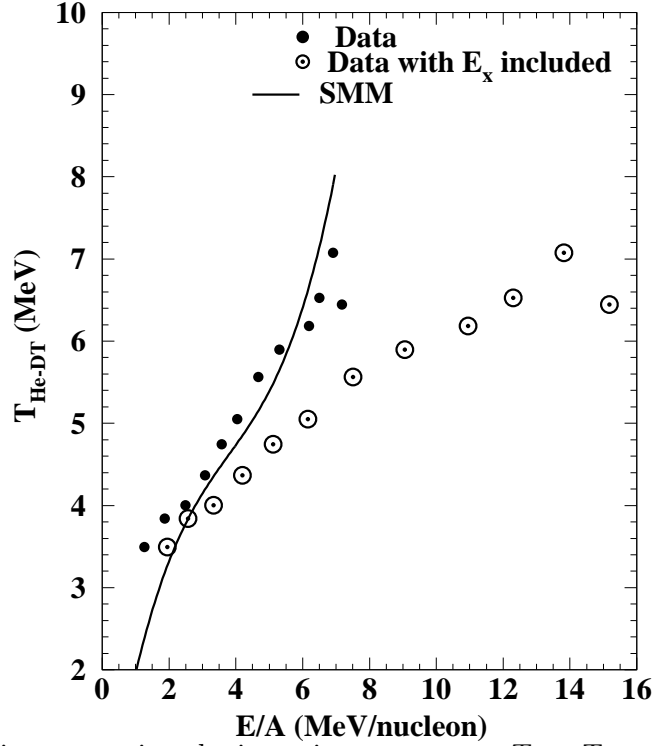}}
\vspace*{0.003in}
\caption{The caloric curve using the isotopic temperature  $T_{a} =
\rm T_{HeDT}$. 
 Experimental
results are shown as a function  of both the input energy/nucleon and
the total excitation energy/nucleon.  The SMM results are for the input
excitation energy/nucleon.
}
\label{fig.13}
\end{figure}
\subsection{Critical exponents and related quantities}

The EOS collaboration has analyzed the 1A GeV $^{197}$Au data in
terms of the theory of critical phenomena, according to which MF is
viewed as a continuous phase transition.  In this section we subject the
SMM events to this same analysis and compare the results with those
obtained from the data.

\subsubsection{The critical point  multiplicity $m_{c}$ 
and the exponent $\tau$}

\begin{figure}
\epsfxsize=8.5cm
\centerline{\epsfbox{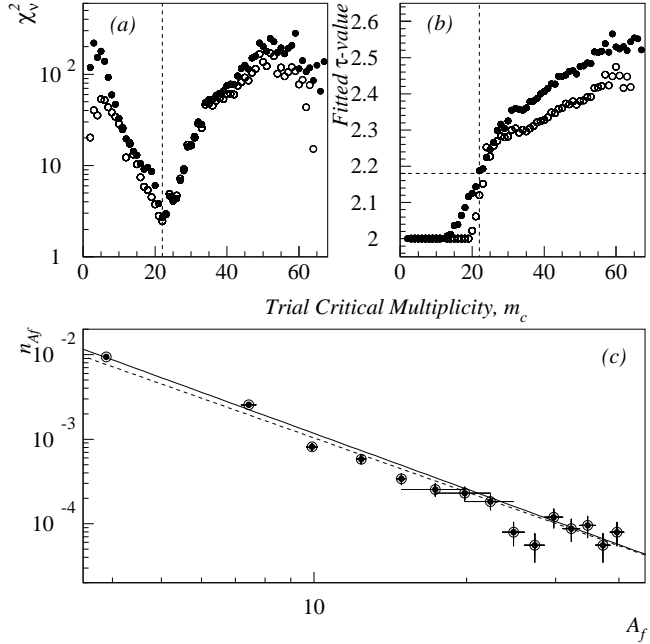}}
\vspace*{0.003in}
\caption{Determination of $\tau$ and $m_c$ from data
\protect{\cite{elliott98}}.  (a) $\chi_{\nu}^2$ values from the power law
fit to the fragment mass yield distribution obtained for different $m$.
Values of $\tau$ as a function of $m$.  (c) Power law fit to data point
$m=m_c$, corresponding to the minimum value of
$\chi_{\nu}^2$. The dashed  line 
is a fit to the open points which exclude
A = 4 fragments.
The block dot results include the A = 4 fragments. 
}
\label{fig.14}
\end{figure}

In order to extract the various critical exponents from the data the location
of the critical point, which can be characterized by the critical multiplicity,
$m_c$, must be determined.  We have  used the method
presented in reference
\cite{elliott96,elliott98}, in which $m_c$, the critical exponents $\tau , \gamma$ and
$\sigma$, and the scaling function were obtained from the EOS data.  We
briefly summarize the procedure below.
The fragment mass yield distribution, $N_{A_f}(\epsilon )$, where
$\epsilon=(m_c -m)/m_c$, is normalized to the mass of the remnant,
$A_{rem}(\epsilon )$.  The normalized fragment distribution can be
written as \cite{stauffer92}
\begin{equation}
n_{A_f}(\epsilon )~=~N_{A_f}(\epsilon )/A_{rem}(\epsilon )~=~q_o
A_f^{-\tau}f(z)
\end{equation}
where $f(z)$ is the scaling function and the scaling variable $z=\epsilon
A_f^{\sigma}$.  {\it If we
assume that scaling is valid for clusters of all sizes},  then $q_o$ is a function of only
$\tau$:
\begin{equation}
q_o~=~1/\zeta (\tau -1)
\end{equation}
where $\zeta$ is the Riemann zeta function and $2<\tau <3$
\cite{nakanishi80,stauffer79}.  At the critical point $f(z)=1$ and a pure
power law is obtained for the fragment mass yield distribution.  The
power law behavior is modified by the scaling function away from the
critical point.
\begin{figure}
\epsfxsize=8.5cm
\centerline{\epsfbox{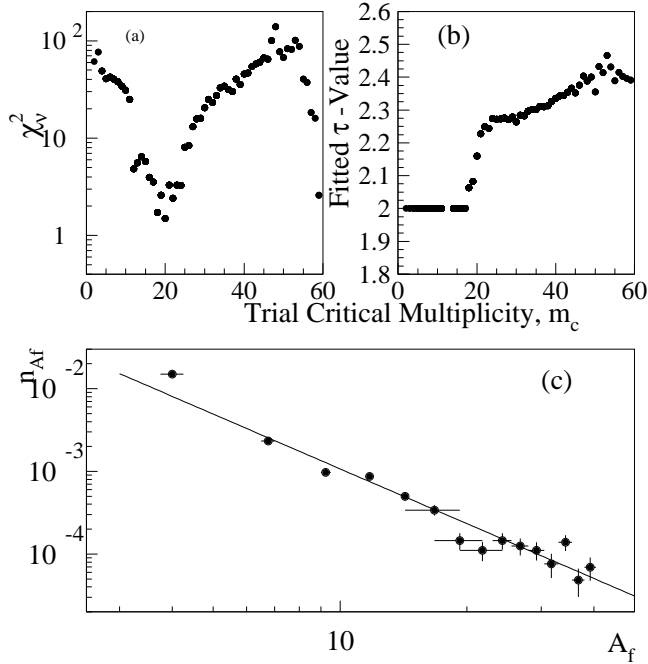}}
\vspace*{0.003in}
\caption{Determination of $\tau$ and $m_c$ from SMM.  See Fig. 14 for details.
}
\label{fig.15}
\end{figure}
In order to determine $m_c$ and $\tau$ one performs power law fits to
$n_{A_f}(\epsilon)$ versus $A_f$ for each value of $m$ over a particular
range of fragments, e.g. $Z_f=3-16$.  The best fit, as determined by the
minimum value of $\chi_{\nu}^2$, gives the value of $m=m_c$.  The
results of this analysis for the EOS data \cite{elliott98} are shown in Fig.
14.  The fitting procedure yields a deep minimum in $\chi_{\nu}^2$
at $m=m_c =22\pm 1$, and thus accurately determines the location of
the critical point.  The power law fit to the data at this multiplicity is also
shown.  The value of $\tau$ is 2.19$\pm$0.02.

The above analysis was also performed for the SMM events for 
$Z_{f} = 3-16$ and the results
are shown in Fig. 15.  The dependence of $\chi_{\nu}^2$ and $\tau$
on $m$ and the quality of the power law fit at $m_c$ are similar to that
exhibited by the data.  The results are summarized in Table \ref{tab2} and
are in excellent agreement with the data.

\subsubsection{Determination of the exponent $\sigma$}

Once $m_c$ is known, it is possible to determine the value of $\sigma$. 
We have used the percolation procedure where the largest piece $Z_{max}, \;
A_{max}$ is removed only  on the liquid side \cite{stauffer92}. 

Since the scaling function must have a single maximum \cite{stauffer92},
we define $z=z_{max}=constant$ as the value of the scaling variable
where the maximum for this fragment mass occurs.  The previously given relation between $z$
and $\sigma$ can then be written for this maximum as
\begin{equation}
z_{max} ~=~\epsilon_{max} A_f^{\sigma}~=~constant
\end{equation}
where $\epsilon_{max}=(m_c -m_{max}(A_f ))/m_c$, with
$m_{max}(A_f)$ being the multiplicity for which the maximum yield of
fragments of mass $A_f$ is obtained.  Eq. (7) leads to a power law,
$\epsilon_{max}\propto A_f^{-\sigma}$, from which $\sigma$ can be
found.

Applying this analysis to SMM , we first determine the values
of $m_{max}(A_f)$.  
Typical results have been shown in Figs. 8
and 9.  Fig. 16 shows the power law plot,
$ln(m_{max}-m_c )$ versus $ln(A_f )$, from which we obtain 
$\sigma = 0.63 \pm 0.08$.  
Applying the same analysis to EOS data, we obtain $\sigma = 0.32 \pm 0.05$,
much smaller than the SMM value\cite{elliottobe}. 
The analysis in Sec. V
suggests that the $\sigma$
values in data could be affected by the hot fragment cooling process, and
that the $\sigma$ value adjusted for cooling  would be $\sigma = 0.54 \pm .11$ 
(See Table~II) This value is used in constructing the scaling function for the
EOS data.
\begin{figure}
\epsfxsize=8.5cm
\centerline{\epsfbox{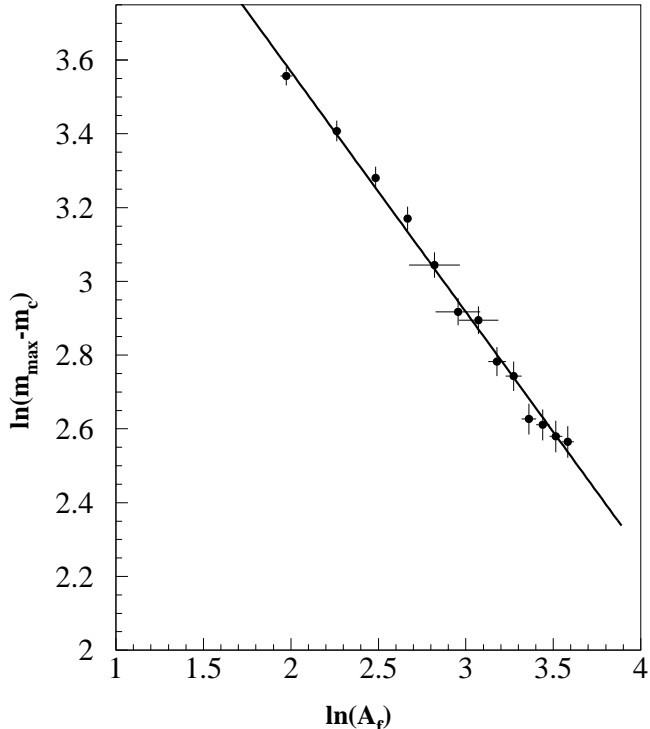}}
\vspace*{0.003in}
\caption{Determination of $\sigma$ from SMM. A linear fit of
ln$(m_{max}-m_c$) vs $ln(A_f)$ for Z=3-15 fragments gives $\sigma = 0.63
\pm .08$. 
}
\label{fig.16}
\end{figure}

\subsubsection{The scaling function}

Knowing the values of $m_c ,~\tau$, and $\sigma$ 
for data and SMM,
it is possible to
evaluate the scaling function by rewriting Eq. (5) as
\begin{equation}
f(z)~=~n_{A_f}(\epsilon)/q_o A_f^{-\tau}.
\end{equation}
The unscaled data is shown in Fig. 17.
Scaling $n_{A_f}(\epsilon)$ according to Eq. (8) collapses the
multifragmentation data from a broad range of fragments yields  onto a 
narrower band
for both data and SMM, as shown in Figs. 18 and 19.  Both scaling functions
show a comparable scatter of the fragments. 
By definition, both have a value of
unity at the critical point, and both have a maximum of comparable
magnitude for virtually the same value of the scaling variable.  
\begin{figure}
\epsfxsize=8.5cm
\centerline{\epsfbox{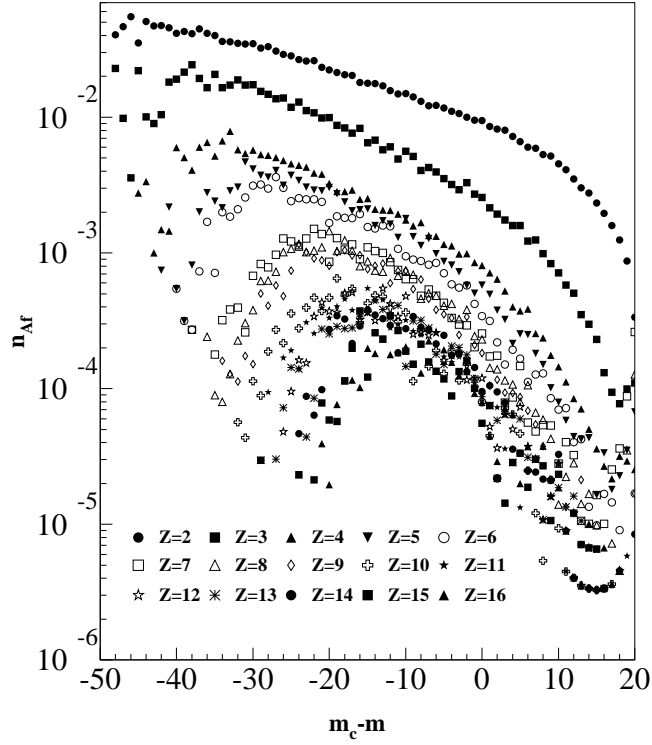}}
\vspace*{0.003in}
\caption{Unscaled experimental fragment yields for $2 \leq Z \leq 16$.
}
\label{fig.17}
\end{figure}
\begin{figure}
\epsfxsize=8.5cm
\centerline{\epsfbox{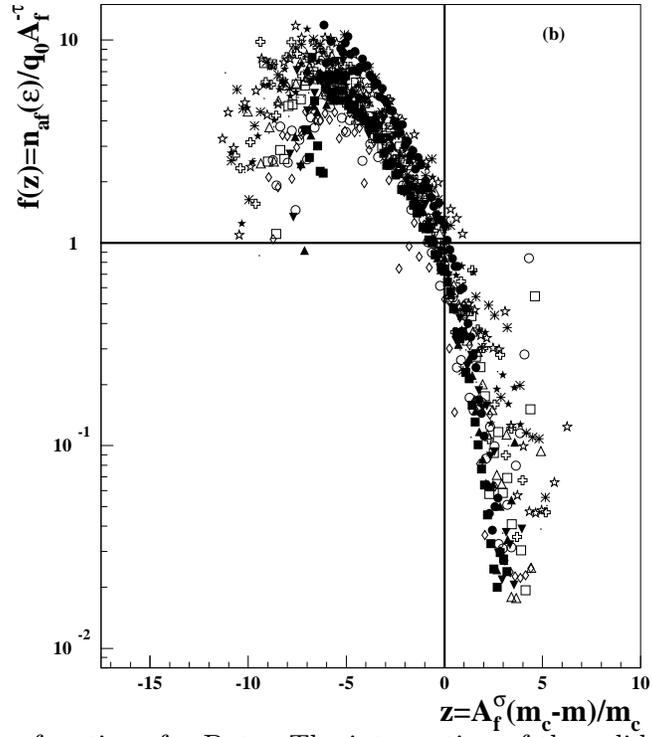}}
\vspace*{0.003in}
\caption{The scaling functions for Data.
 The intersection of the solid lines marks the critical point. 
}
\label{fig.18}
\end{figure}

\subsubsection{The exponent $\gamma$ and moments of the fragment
yield distribution}

The exponent $\gamma$ has been determined by means of the
$\gamma$-matching technique as applied to the second moment, $M_2
(\epsilon )$, of the fragment yield distribution \cite{gilkes94}.  The second
moment is defined as
\begin{equation}
M_2 (\epsilon )~=~\Sigma n_{A_f}(\epsilon )A_f^2
\end{equation}
and a similar expression may be used if the fragments are
characterized by their nuclear charge instead of mass.  Again,
following the percolation theory procedure,  we omit the largest
fragment from the summation in Eq. (9) only on the ``liquid'' side of the
critical point $(m<m_c )$.  All fragments are included on the ``gas''
side $(m>m_c )$.  The exponent $\gamma$ is obtained from the power law
$M_2 \propto |\epsilon |^{-\gamma}$ by searching for various multiplicity
regions that yield values of $\gamma_{gas}$ and $\gamma_{liquid}$ that
agree with each other.  The value of $m_c$ is again determined in the same
fit.
\begin{figure}
\epsfxsize=8.5cm
\centerline{\epsfbox{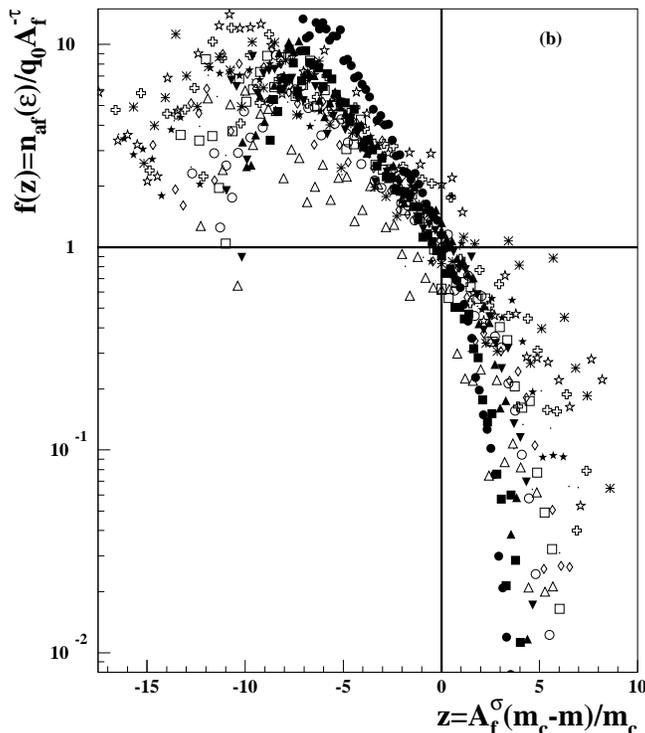}}
\vspace*{0.003in}
\caption{The scaling functions for SMM.
 The intersection of the solid lines marks the critical point
}
\label{fig.19}
\end{figure}
The results for SMM are shown in Fig.20.
The procedure has also  been described in Ref. \cite{elliott96}
and the results obtained from an analysis of the EOS data are
summarized in Table \ref{tab2}.  The same analysis can be applied to
the SMM events  and  the results are  also
summarized in Table \ref{tab2}.  The experimental and calculated
$\gamma$ values are consistent within the limits of error.

We have also evaluated several other quantities that have been associated
with critical behavior \cite{campi86,campi88,srriv99}.  They include the
fluctuations in the size of the largest fragment and in $M_2$,  the
magnitude of the peak in the combination of moments $\gamma_2 = M_2
M_0 /M_1^2$, and the determination of $\tau$ from a plot of ln$M_3$
versus ln$M_2$.  These quantities, too, are in good agreement with
experiment.  
\begin{figure}
\epsfxsize=8.5cm
\centerline{\epsfbox{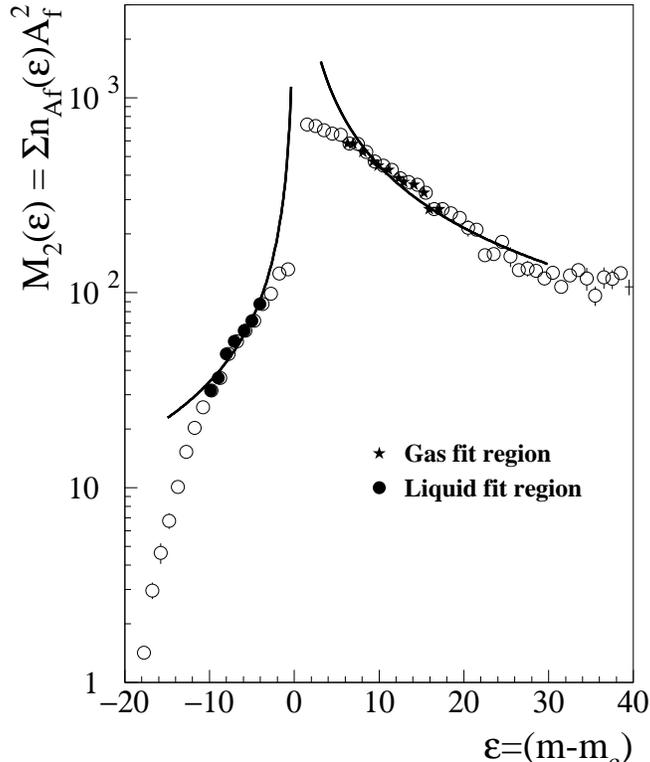}}
\vspace*{0.003in}
\caption{Second moment of the SMM generated fragment  mass yield distribution.The largest fragment has been removed only from the ``liquid'' side $(m<m_c)$.All points are from SMM. The solid points from SMM  have been used to 
determine $\gamma$. The solid curve is the fit of the power law 
$M_{2} = C_{\pm} | \epsilon |^{-\gamma}$. 
}
\label{fig.20}
\end{figure}
The fluctuations of the size of the largest fragment and of the moments
provide an independent method to locate the critical multiplicity $m_{c}$
of the  MF transition. The first moment $M_{1}$ can also indicate how the nucleons are
distributed into light particles, IMFs and the largest piece in MF.
This distribution can identify the phases present in MF and if
the coexistence of liquid and gas phases occurs in MF. (See Sec.\ VI.)

\section{EFFECTS OF COOLING AND OF THE USE OF A SINGLE REMNANT}

The SMM calculations described in the preceding sections give results for
deexcited secondary fragments formed from a distribution of remnants. 
These remnants are grouped according to multiplicity and we have noted
that at a given $m$ there is a distribution of remnant charges, masses, and
excitation energies.  An ideal statistical analysis of multifragmentation
would involve primary fragments formed from the breakup of a unique 
remnant $(A,Z)$ as a function of remnant excitation energy or temperature.  In this
section we use SMM to examine the extent to which the departures from
this ideal situation affect the values of the critical exponents and related
fragment properties.  We introduce the terms SMM$_{hot}$ and
SMM$_{cold}$ to designate results obtained from SMM for primary
fragments and deexcited fragments, respectively.

\subsection{Hot and cold SMM fragment yields}

The most direct view of the effect of deexcitation is provided by a
comparison of the SMM$_{hot}$ and SMM$_{cold}$ fragment yield
distributions.  For simplicity, we evaluate these distributions for the
decay of a single remnant as a function of $E_i$.  We have chosen the
remnant formed in the Au on C interaction corresponding to the critical
multiplicity in the data, $A = 160, \; Z = 64$, and evaluated its
breakup for a range of $E_i$ values. At a given $\it m $ the total rms width
of the remnant distribution is $\sim $ 5-7 \%.
 In SMM the average Z/A ratio of the hot
``pre-fragments'' is the same as the remnant Z/A ratio,
i.e., no particles or energy leaves the remnant system during the $\sim$
 100 fm/c time frame for MF.
\begin{figure}
\epsfxsize=8.5cm
\centerline{\epsfbox{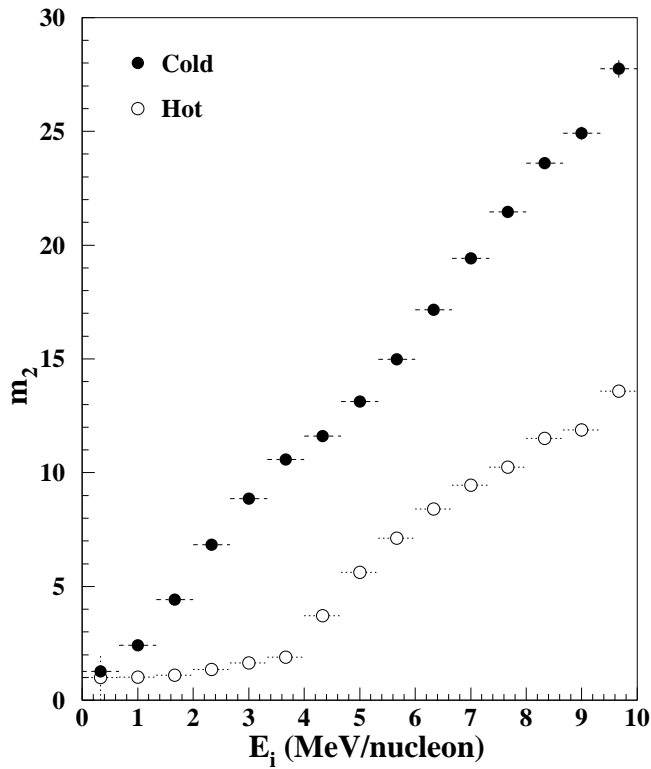}}
\vspace*{0.003in}
\caption{Second stage multiplicity $m_{2}$ for SMM$_{cold}$ and SMM$_{hot}$.
}
\label{fig.21}
\end{figure}
The increase in the charged particle multiplicity $m_{2}$ from SMM$_{hot}$ to
SMM$_{cold}$ is shown in Fig.~21. This figure shows that cooling is
important.
Typical results for fragments with $Z>3$ are shown in Figs. 22
and  23.  The fragment multiplicities for SMM$_{hot}$ and
SMM$_{cold}$ are in close agreement up to $E_i ~\simeq~7$ MeV/nucleon.
However, at higher excitation energies the primary fragment yields are
significantly larger than those of the corresponding cold fragments.
At these high $E_i$, fragments undergo substantial secondary decay,
which according to SMM, occurs by evaporation for fragments with $A > 16$.
 For $A<16$, secondary decay occurs  by Fermi
breakup.  These mechanisms tend to form the lightest particles and
thereby reduce the yields of fragments with $Z\geq$3.  We emphasize the
fact that the net effect of these secondary processes 
on the fragment distributions 
is rather small below 7
MeV/nucleon, the energy range where most of  the MF occurs.  The
critical point corresponds to a much lower excitation energy, $E_i \sim
4.3$ MeV/nucleon \cite{hauger98}, and therefore quantities determined
at these low energies should be nearly unaffected.
\begin{figure}
\epsfxsize=8.5cm
\centerline{\epsfbox{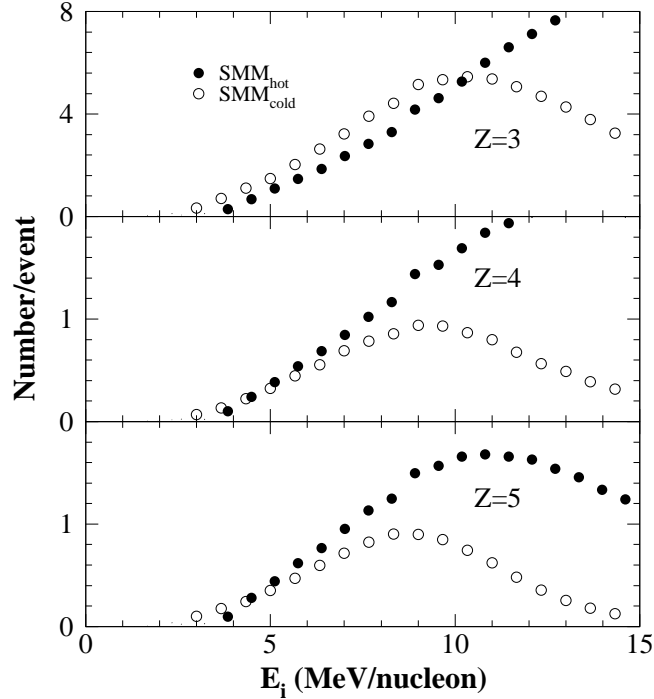}}
\vspace*{0.003in}
\caption{Multiplicity of Z=3, 4, 5 fragments as a function of input
energy for SMM$_{hot}$ and SMM$_{cold}$ in MF of A = 160, Z = 64.
}
\label{fig.22}
\end{figure}
Fig. 24 shows the results for Z=1 and Z=2 particles.  Contrary
to the results shown in Figs. 22 and 23, most of these
particles are formed in the cooling step.  Averaging over the
excitation energy range important for MF, 3-7 MeV/nucleon, we find that
the lightest hot fragments have a higher internal energy per nucleon
than the heavier hot fragments, $\sim$4 MeV/nucleon versus $\sim$3
MeV/nucleon.  Consequently the main source of these particles is the
Fermi breakup of the lighter fragments.

\begin{figure}
\epsfxsize=8.5cm
\centerline{\epsfbox{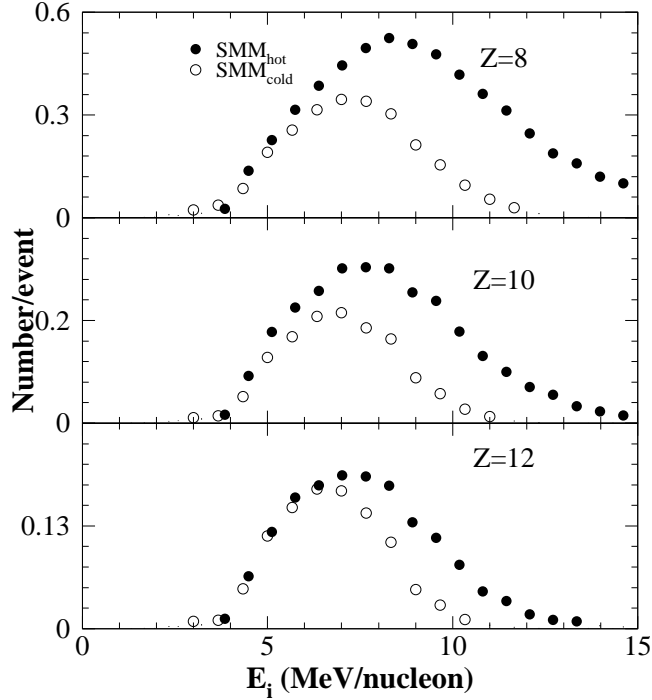}}
\vspace*{0.003in}
\caption{Multiplicity of Z=8, 10, 12 fragments as a function of input
energy for SMM$_{hot}$ and SMM$_{cold}$ in MF of A = 160, Z = 64. 
}
\label{fig.23}
\end{figure}

\subsection{Effect of cooling on critical exponents and related
quantities}

The analysis described in Section IVC has been applied to the products of
SMM$_{hot}$, using the experimental remnants as input.  The resulting
values of the various exponents are tabulated in Table \ref{tab3}.  The
calculated value of $\tau$ for hot fragments is $\sim 6\%$ smaller than
that for cold fragments while that of $\gamma$ is $\sim 11\%$ smaller. 
 In contrast, the determination of the
 exponent $\sigma$ differs significantly for the two distributions,
with the SMM$_{hot}$ value being $\sim$40\% larger than that from
SMM$_{cold}$.  The large difference in $\sigma$ values is understandable
given that $\sigma$ is determined in part by the multiplicities for which the various
fragments attain their highest yields.  As shown in Figs. 8 and
9, the yields of the lighter fragments peak at large values of $m$,
corresponding to $E_i$ values of 7 MeV/nucleon and higher.  It is precisely
at these energies that secondary decay reduces the yield of these fragments
and thereby leads to a smaller $\sigma$. 

\begin{figure}
\epsfxsize=8.5cm
\centerline{\epsfbox{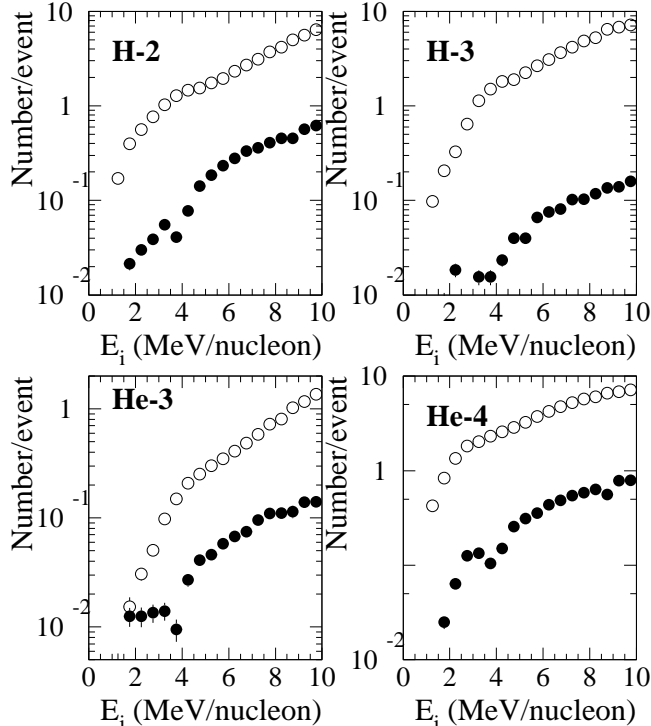}}
\vspace*{0.003in}
\caption{Multiplicity of Z=1 and 2 particles as a function of input
energy for SMM$_{hot}$ (solid circles) and SMM$_{cold}$ (open circles)
in MF of  A = 160, Z = 64. Note that most of these particles are produced in the cooling process.
}
\label{fig.24}
\end{figure}

In a recent publication Mastinu et al. \cite{mastinu98} used the
Copenhagen model to examine the effect of secondary fragment decay in
the MF of $^{197}$Au.  In particular, they compared the shape of the
second moment, $M_2$, of the fragment charge distribution for
SMM$_{hot}$ and SMM$_{cold}$.  They found that the shape of the $M_2$
distribution in the vicinity of its peak is unaffected by secondary
decay when plotted against $E^*$ whereas there is a change in shape
when $M_2$ is plotted against $m$. 

 On this basis they concluded that
secondary decay affects the values of exponents such as $\gamma$, which
is obtained from $M_2$, when charged particle multiplicity is used as
the control parameter.  Unfortunately, they did not calculate the
magnitude of this effect.  We have already evaluated the effect on
$\gamma$ of secondary decay and, as shown in Table \ref{tab3}, find
that it is a relatively small 10$\%$ correction.  We note in passing
that the $M_2$ plot shown by Mastinu et al. \cite{mastinu98} is not appropriate for the
extraction of $\gamma$ because the largest fragment has been removed
from {\it both} the liquid and gas sides instead of only on the liquid side,
as was done in Fig.~20.  Furthermore, they did not turn off
the fission channel in their SMM calculation and the contribution of
this mechanism disproportionally affects the analysis.

\subsection{Critical exponents from the multifragmentation of a single
remnant for SMM$_{\bf hot}$}

The analysis described in Section IVC is applied to the breakup
of a single remnant, using the {\it input energy} as the control
parameter.  We use the  critical
remnant  A~=~160, Z~=~64  as the system size for SMM.
Fig. 25 shows the
SMM$_{hot}$ results for the determination of the  critical point by
means of the one-parameter power law fit to the fragment mass yield
distribution.

Again, there is a deep minimum in $\chi_{\nu}^2$ which
determines the location of the critical point.
This should be compared with 
the same results for
SMM$_{cold}$ using  the experimental remnant distribution and $m$ as
the control parameter (Fig. 15).
\begin{figure}
\epsfxsize=8.5cm
\centerline{\epsfbox{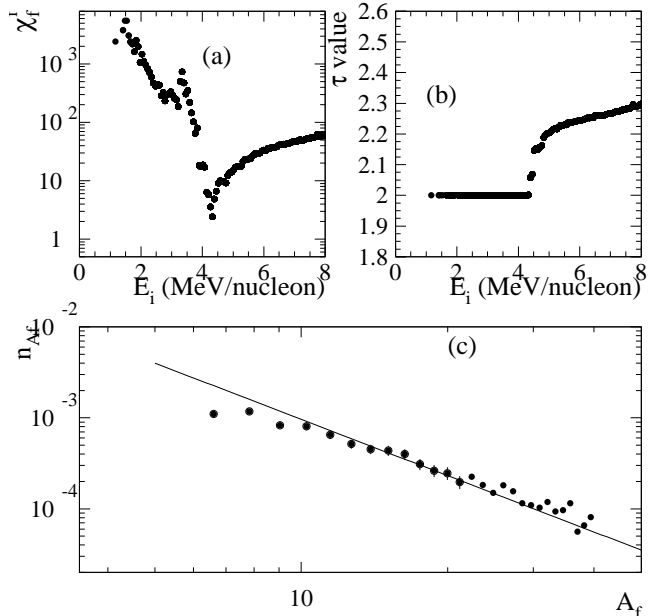}}
\vspace*{0.003in}
\caption{Determination of $\tau$ and $E_{c}$ in the MF of  A = 160, Z = 64
for SMM$_{hot}$.  The panels have the same meaning as in Fig. 14
except that $E_i$ replaces $m$ as the control
parameter.
}
\label{fig.25}
\end{figure}
The value of $\tau$ obtained from the fit in Fig. 25 is given in
Table \ref{tab3}.  Also included are the values of the other
exponents.  A comparison with the corresponding exponents obtained for
SMM$_{cold}$ for the experimental remnant distribution indicates that
the difference in $\tau$ and $\gamma$ values is less than 10$\%$.  Thus
the combined effect of secondary decay, the remnant distribution for a
given $m$,  and the use of
multiplicity as the control parameter have only a minor effect on the
values of the $\tau$ and $\gamma$ exponents.  However, the exponent $\sigma$ 
for SMM$_{hot}$ is much
larger  than the corresponding value of SMM$_{cold}$ just as was the case for the
experimental remnant distribution. Again, this difference reflects the effect
of cooling.
We note here that the use of a single remnant and
excitation energy as the control parameter, which is most readily seen
in a comparison of the two SMM$_{hot}$ columns, has essentially no
effect on the critical exponents.
\begin{figure}
\epsfxsize=8.5cm
\centerline{\epsfbox{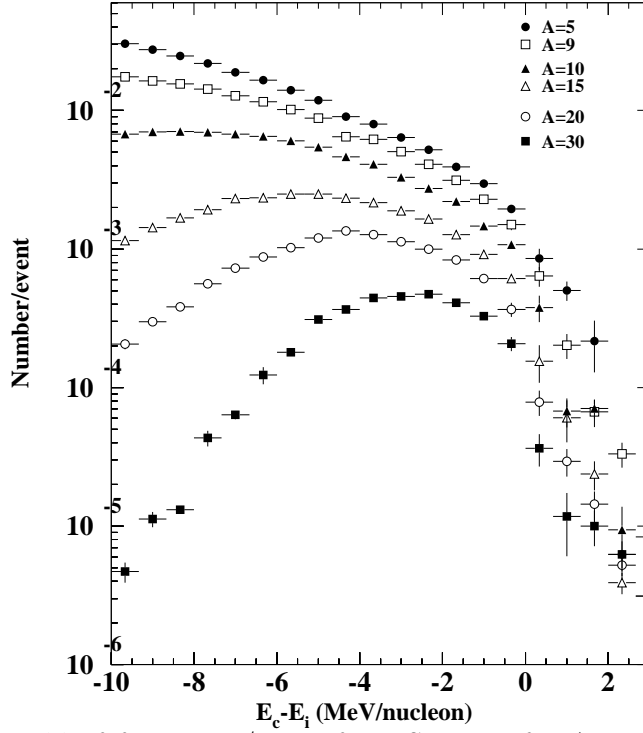}}
\vspace*{0.003in}
\caption{Unscaled yield of fragments/event from SMM$_{hot}$ for A = 5 $-$ 30 as a function of $E_{c} - E_{i}$.
}
\label{fig.26}
\end{figure}
Fig. 26 shows the unscaled SMM$_{hot}$ fragments for  A = 5
to 30 for the single remnant system.
Fig. 27 shows the scaling function obtained from
these data and the resulting 
scaling collapse of the data
into a very narrow band.  
This scaling function should be compared with the scaling function from
SMM$_{cold}$ (Fig.~19).
  The more extensive scatter in
Fig.~19 can be attributed in comparable measure to the effects
of fragment cooling and to the presence of the remnant distribution, 
using multiplicity  as the control parameter.  This indicates that the
scatter seen in the experimental  data at the ends of the scaling curves, 
Fig. 18, may also be due to these effects.
\begin{figure}
\epsfxsize=8.5cm
\centerline{\epsfbox{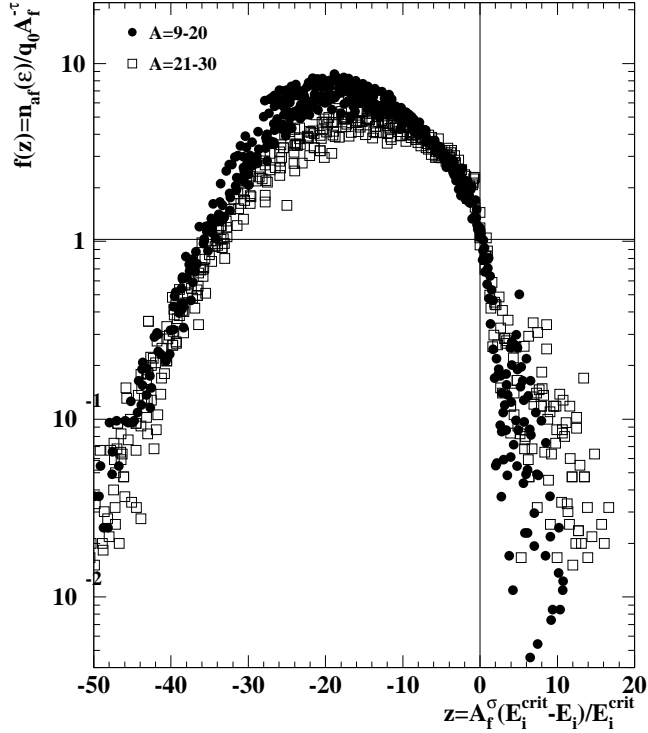}}
\vspace*{0.003in}
\caption{Scaling function obtained from SMM$_{hot}$ for MF of
A = 160, Z = 64.
}
\label{fig.27}
\end{figure}
\section{NATURE OF THE PHASE TRANSITION IN SMM}

We have shown that SMM can reproduce the various features of the EOS
data, including the values of most of the critical exponents, critical
multiplicity, and critical scaling.  If SMM did not predict the
occurrence of a continuous phase transition then the exponents would merely
constitute a particular parameterization of the data and would have no
ulterior significance.  It is therefore important to determine 
if SMM  predicts such  a thermal phase transition. 

We follow the approach of Gross \cite{gross97a} and H\"{u}ller
\cite{huller94}, who show that the nature of the transition can be
determined by means of the microcanonical equation of state,  using 
a plot of the reciprocal partition  temperature 
$\beta_{p} = 1/T_{p}$
versus $E_i$. We assume $T_{p}$ is the best estimate of the average
value of the  fluctuating event-by-event 
SMM microcanonical 
temperature.  A first order transition is identified by
backbending in the $\beta_{p}$ vs $E_{i}$ curve and hence has a region of
negative system specific heat, 
$\displaystyle C_{n} = - \beta_{p}^{2}/ \frac{d\beta_{p}}{dE_{i}}$.  
A continuous phase transition will not
exhibit backbending and will have a positive system specific heat  that peaks
at the critical energy. 
\begin{figure}
\epsfxsize=8.5cm
\centerline{\epsfbox{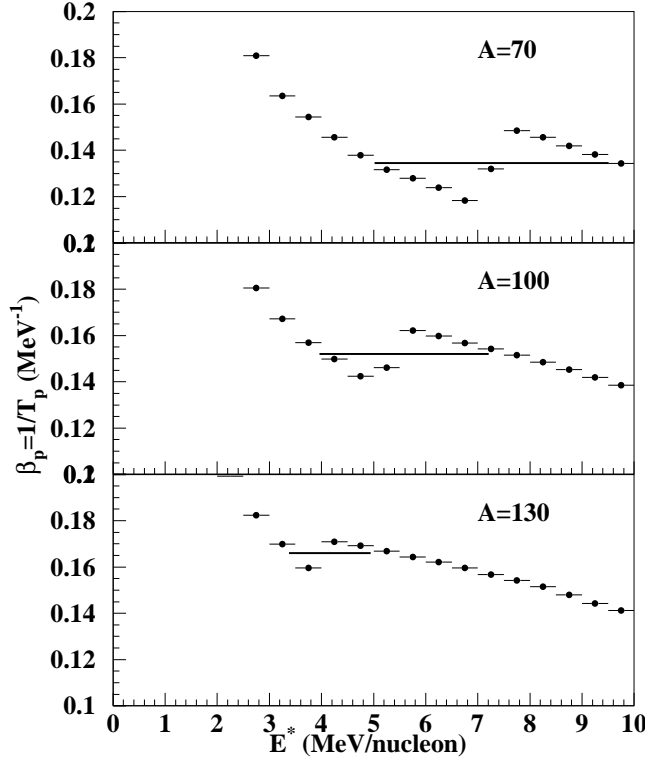}}
\vspace*{0.003in}
\caption{Dependence of $\beta_{p}$
on the input energy for A=70, 100 and 130. The lines are drawn for
equal area  and show the  ``Maxwell Construction''.
}
\label{fig.28}
\end{figure}

Recall that for  infinite neutral nuclear matter a critical phase
transition is expected at a temperature $\sim$15--18 MeV
\cite{ravenhall83}.  At lower temperatures, infinite 
neutral nuclear matter would exhibit a first order liquid-gas phase
transition  with possible coexistence of constant density phases.
SMM, however, describes the transition of a highly excited
{\it finite  charged} nuclear drop into a number of hot intermediate
mass fragments followed by a de-excitation process. The cold state
liquid and gas 
consists of smaller droplets and nucleons emitted in the cooling process. 
In the hot 
stage, the so-called pre-fragments 
are produced in thermodynamic
equilibrium where the volume, surface, Coulomb,
symmetry and translational 
terms determine the fragment yields \cite{bondorf95}.

To investigate the interplay of these energy terms we study SMM$_{hot}$
as a function of the remnant mass and charge \cite{schar99}.
 Fig.28 shows $\beta_{p}$ versus $E_i$ curves for A~=~70, Z~=~30;
 A = 100, Z = 40; and A  = 130, Z = 53. A characteristic
backbending is observed for these systems.  Following the
microcanonical prescription, we make the ``Maxwell construction''
\cite{gross97a}, which determines the  average transition temperature
and the magnitude of the transition energy or effective latent heat.  
For $A$=100 we obtain a
transition energy of 3.2 MeV/nucleon
as given by the width of the backbend region.
  Figure 29 shows that the
system specific heat for A = 100 is negative indicating the possibility of a first order
phase transition. We have computed  the effect of the variable volume
on the system specific heat and found that it is a very small fraction
of the total.  The translational energy contribution is also very small.  Thus
$C_{n}$ describes the change in the internal energy of the MF system.
The transition energy plays the role of an effective  latent heat for
MF. 
SMM$_{hot}$ indicates that the effective latent heat is reduced for heavier
remnants.
For the A = 70, 100, and 130 systems, there is a
progressive decrease in the transition energy as shown in Fig.30. 
A linear extrapolation suggests that the transition energy
as an effective latent heat  could vanish for A $>$ 170. 
We can gain further insight about these transitions by studying the
distributions of light particles and fragments prior to the
de-excitation or cooling process. For A = 100
 the first moment of the hot
mass distribution as a function of the input energy is shown in Fig.
31.  For $E_{i} = 8$ MeV/nucleon  only $\sim$ 5 out of 100
nucleons are not part of the A$>$4  droplet distribution.  Thus hot
SMM describes a transition where a large droplet is multifragmented
into smaller droplets. The subsequent deexcitation processes are
responsible for most of the light particles with A $<$5 seen in the
asymptotic mass distributions. 
\begin{figure}
\epsfxsize=8.5cm
\centerline{\epsfbox{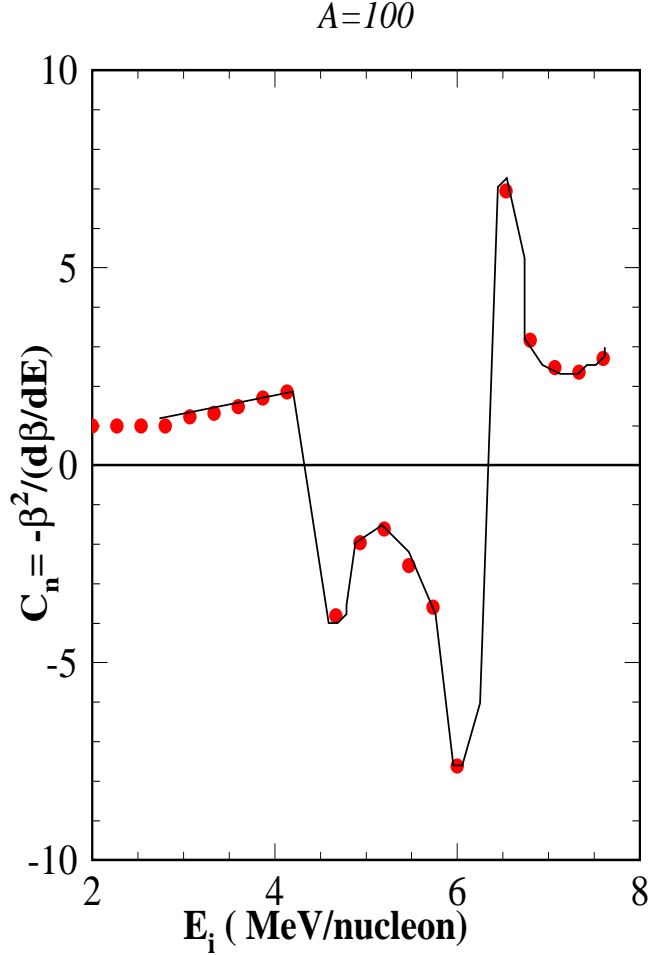}}
\vspace*{0.003in}
\caption{System specific heat per particle for A=100. The solid curve
is to guide the eye. 
}
\label{fig.29}
\end{figure}

In this SMM model backbending can be a relatively large
effect. The extra surface energy
required for the MF transition cools the system and the temperature falls.
For a larger system the fractional change in the surface to volume ratio
due to MF is reduced and the backbending is smaller. In SMM the total Coulomb 
energy is reduced by the factor $ 1 - 1/(1+\kappa)^{1/3} \sim 0.3$
due to expansion  prior to
clusterization. Since the total energy is measured with respect to the ground
state of the  remnant system 
$A$, $Z$ at normal density, this introduces a $Z$  
dependence into the $\beta_{p}$ versus $E_{i}$ equation of
state.

Fig. 30b shows the average transition  temperature, which
decreases as the remnant mass and charge  increases. This decrease  
clearly reflects the increase
in Coulomb energy for the heavier remnants \cite{levit85}.  In contrast, 
for a
liquid-gas phase transition in {\it finite neutral matter} we would expect the
transition temperature to increase with the system
size, because the binding
energy per nucleon increases. 
For a larger charged nuclear remnant, 
the transition energy will vanish  if enough Coulomb energy is 
converted to heat energy in the expansion of the remnant prior to the
 MF transition. 
\begin{figure}
\epsfxsize=8.5cm
\centerline{\epsfbox{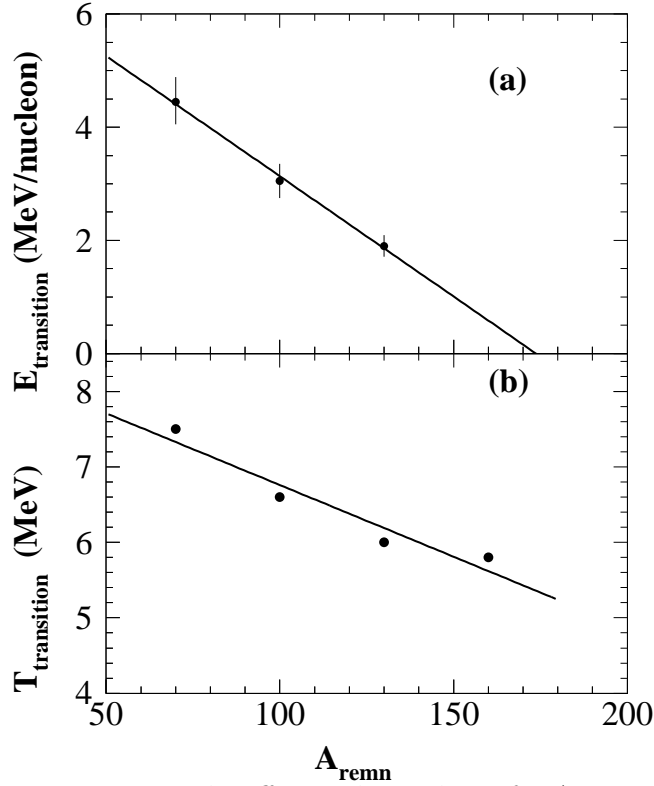}}
\vspace*{0.003in}
\caption{a)  
MF transition energy, the effective latent heat, for A = 70, A = 100,
and A = 130. 
b) Average, i.e., Maxwell construction,  transition temperature for 
A = 70, A = 100, A~=~130, and A~=~160.
}
\label{fig.30}
\end{figure}

For the remnant mass $A$ = 160 the reciprocal partition temperature $\beta_{p}$ is plotted versus $E_i$ in Fig. 32a
and the fluctuating  event-by-event reciprocal microcanonical  temperatures
$\beta_{p}$, are shown in Fig. 32b.
 The area under the $\beta_{p}$ vs $E_i$ curve is  the system entropy.
In Fig. 33a, we expand the $\beta_{p}$ vs $E_{i}$
plot. The system specific heat per particle is computed 
by taking differences between adjacent points. $C_{n}$ is shown in Fig. 33b.
The specific heat is positive 
and peaks near the critical energy,
indicating a continuous phase transition.  
H\"uller characterizes such behavior as a continuous phase
transition with an anomalous specific heat!\cite{huller94}
\begin{figure}
\epsfxsize=8.5cm
\centerline{\epsfbox{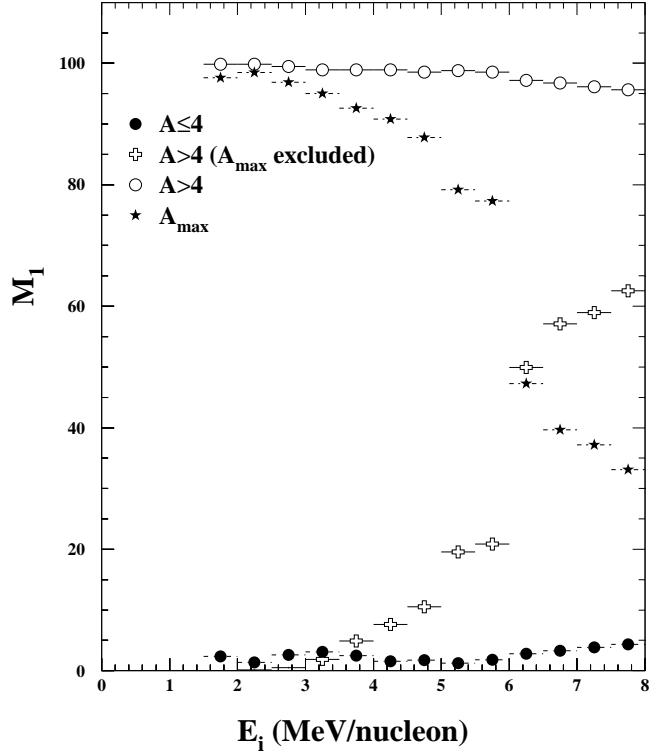}}
\vspace*{0.003in}
\caption{First moment of the $SMM_{hot}$ mass distribution for A = 100.
}
\label{fig.31}
\end{figure}
A method to determine the order of the phase transition based on
canonical kinetic energy fluctuations has recently been
proposed\cite{agos99,chomaz99,agos00}.  To test the conclusions based
on the SMM$_{hot}$ microcanonical equation of state, we have done this
type of analysis on the Au on C data.  For a  single remnant system
size $A,Z$ this analysis compares the canonical specific heat $C_{1}$
with the thermally scaled kinetic energy fluctuations  $(E^{2} - < E >
^{2}) / T^{2}$.

  Following the prescription used on the 35 MeV/nucleon
Au on Au data we find that the energy  fluctuations from the Au on C
data are dramatically reduced when a $\Delta A/ A \leq  0.03$ remnant
mass cut is made, which assures that we have a single system size, the
necessary constraint for the fluctuation analysis.  Thus for the Au on
C data, we find that  $C_{1} > (E^{2} - <E>^{2}) /T^{2}$ in the whole
MF region. The value of $C_{1}$ still remains positive even for a
larger remnant mass cut e. g. $\Delta A/ A$ $\sim$ 10\%. This result argues for a continuous phase
transition. A report of this work is in preparation.

  Very recently,(since the submission of the present article) theoretical
analyses of the nature of MF phase transition and its effect on the caloric curve have been published \cite{chomez00,mulken00}, indicating the occurrence of a
first order phase 
transition. Both of these calculations neglect the Coulomb interaction, which has been shown 
to play an important role in MF \cite{levit85}. It is obvious that there will 
be a first order phase transition in absence of Coulomb energy.

\begin{figure}
\epsfxsize=8.5cm
\centerline{\epsfbox{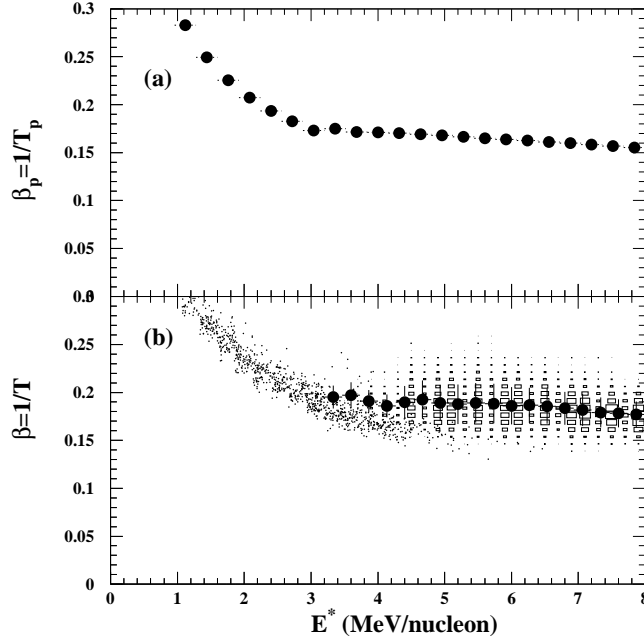}}
\vspace*{0.003in}
\caption{a). Dependence of $\beta_{p}$
 on the input energy $E_{i}$
 for A = 160.  b). Event to event fluctuations of 
$\beta$ 
versus the input energy for A = 160. The dot scatter points  identify
compound nucleus events, the box scatter points identify MF events. The
solid block dots are the average $\beta$ values for MF events.
}
\label{fig.32}
\end{figure}

It has been suggested on the basis of a
caloric curve using the 
 SMM partition temperature
versus an experimentally determined 
 input energy, that MF of gold is a first order liquid-gas phase
transition \cite{poch95}. 
This curve, computed for comparison with
the ALADIN data 
showed that the temperature first
increases with excitation energy, then remains nearly constant at 5-6
MeV as $E_i$ increases between 3 and 10 MeV/nucleon, and then again
increases.  The first regime has been interpreted as the liquid phase,
i.e. the compound nucleus, the second as the MF coexistence phase, and
the third as the gas phase, consisting of a mixture of nucleons and a
few of the lightest fragments \cite{bondorf98}.  
This viewpoint can be probed by examining the first moment of
the {\it hot fragment} yield distribution
for a heavy remnant. Fig. 34a shows a plot of
the dependence on $E_i$ of the first moment of the hot fragment yield
distribution for the experimental remnant distribution.
\begin{figure}
\epsfxsize=8.5cm
\centerline{\epsfbox{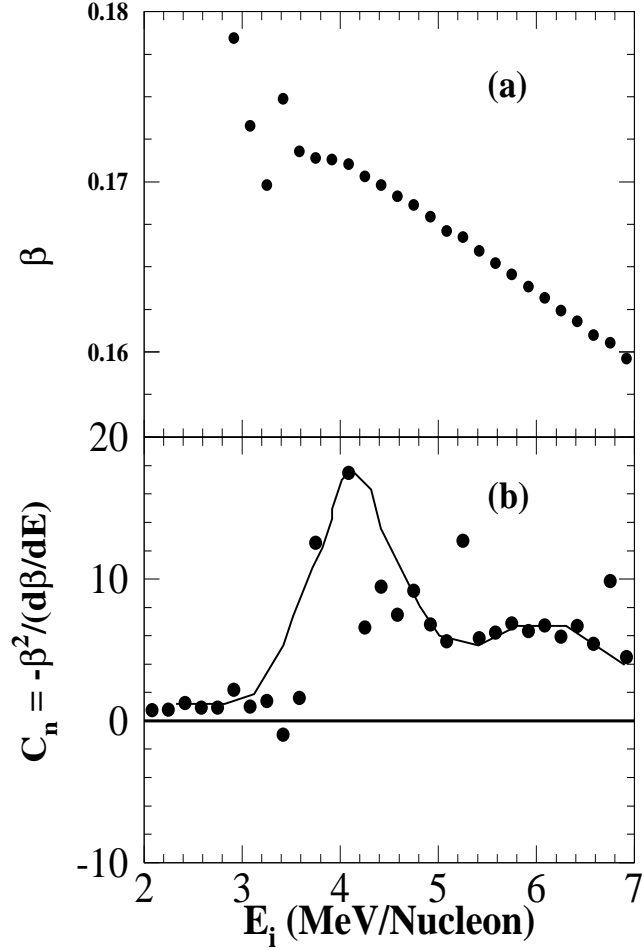}}
\vspace*{0.003in}
\caption{a). Expanded $\beta_{p}$ vs $E_{i}$ curve.
b). System specific heat per particle for A=160.
}
\label{fig.33}
\end{figure}
The figure shows
that even for $E_i \sim 8$ MeV/nucleon, well above the MF transition,
 only about 8\% of the remnant mass ends up in particles with $ A\leq 4$
and that the other 92\% are in intermediate mass fragments with $A>$4.
This argues against the coexistence of the constant density
liquid and gas phases in SMM.
For the experimental remnant distribution,
we compare the
first moment $M_{1}$, for SMM$_{hot}$ Fig. 34a with $M_{1}$ for SMM$_{cold}$
as shown in Fig. 34b. 
The cooling of the hot fragments which carry the multifragmentation
signal produces a large number of final state nucleons and light composites
in agreement with the data.
The IMFs survive the cooling process and identify
the MF transition. This is due in part to the low $\sim$ 4.3 MeV/nucleon
critical energy  for MF. For an A = 70, Z = 30 remnant , 
the predicted center of the backbend region energy is $\sim$
7.5 MeV/nucleon. Here, the hot fragments will have to emit a significantly higher
fraction of their fragment mass in the cooling process and the MF signal could
be severely attenuated. Thus the  Coulomb energy, which lowers the MF
threshold and reduces the effective latent heat, plays a central role in MF.
 Here the
vanishing 
transition energy reflects the changes in the surface, volume, and Coulomb
energies associated with MF rather than the traditional latent
heat in a liquid-gas phase transition, which primarily involves the transfer of individual nucleons from the liquid to the gas, where both phases are at
constant density $\rho_{liquid}$ and $\rho_{gas}$.
\begin{figure}
\epsfxsize=8.5cm
\centerline{\epsfbox{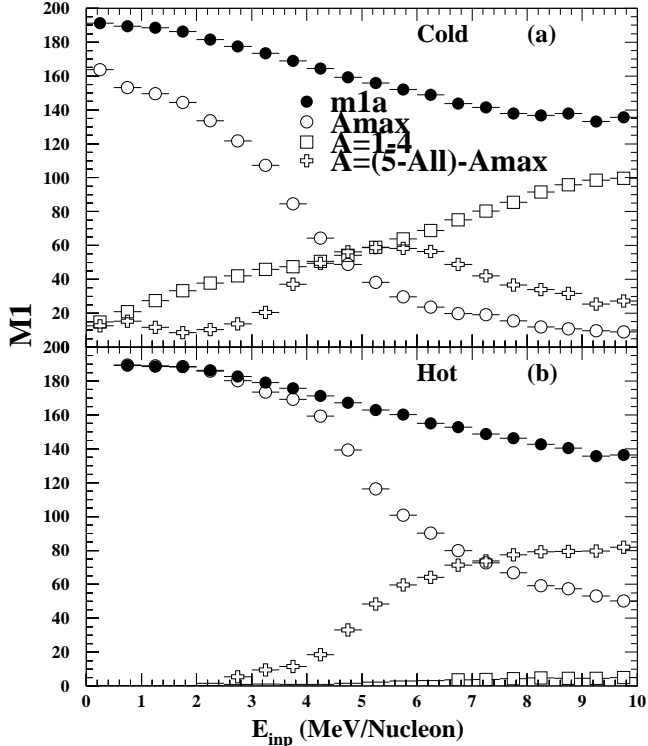}}
\vspace*{0.003in}
\caption{a) First moment of the mass distribution from SMM$_{cold}$ for
the experimental remnant distribution. 
b) First moment of the mass distribution from SMM$_{hot}$. 
The cooling mechanism is nucleon  and $A \leq 4$ composite
particle emission.
}
\label{fig.34}
\end{figure}

\section{SUMMARY AND CONCLUSIONS}

We have compared the EOS multifragmentation results for 1A GeV Au on C
with the SMM model where the volume parameterization is determined by
experiment.  We have found that the standard SMM
parameters produced excellent agreement.  
The input to SMM consisted of
the $Z, A$,   and $E_i$ of the individual remnants produced in some
32,000 fully reconstructed events.  Thus, in contrast to earlier 
experiments which have been compared with SMM, the 
$A, Z, E_{i}$ 
event-by-event input
data were obtained from experiment rather than by using SMM to
constrain average input conditions, or by the use of a theoretical first
stage transport model.

We find that SMM is in very good agreement with the observed fragment
charge yield distribution, fragment multiplicity distributions, total
charged particle multiplicity, isotope ratio temperatures  and 
caloric curve.
This agreement is obtained when the expansion energy is
subtracted from the experimentally determined  excitation energy
of the remnant.  The agreement provides further confirmation that
multifragmentation can be described as an equilibrium process.

SMM also predicts a power law for the fragment mass yield distribution
at essentially the same multiplicity or excitation energy as is
observed in the data.  
Other features characteristic of a continuous phase transition, such as
the scaling function, are also reproduced.  The critical exponents
$\tau$, $\gamma$, and $\sigma$ were obtained from  SMM for comparison
with the experimental values.  The agreement is excellent for $\tau$,
which is based on results obtained just at the critical point, fair for
$\gamma$ 
which is based on results obtained above and below the critical point,
 and  poor for $\sigma$, which is based on results 
well above the critical point.

The effect of cooling of the primary hot fragments produced by SMM has
been evaluated.  The primary hot fragment yield distributions for $Z\geq$3
are only minimally affected by the cooling process below $E_i$ =7
MeV/nucleon.  
The SMM $\sigma$ values are substantially  affected
by cooling, while $\tau$ and $\gamma$ are unaffected by cooling.
The scaling function obtained for
SMM$_{hot}$ is fully collapsed 
into a very narrow band.
In contrast to fragments with Z$\geq$3, the yield of particles with
Z$\leq$2 is substantially increased by cooling, primarily as a result of the
Fermi breakup of light fragments.

We have also performed SMM calculations in which the experimental
remnant distribution has been replaced by a single average remnant.  The
results are virtually unchanged indicating that the mixing of different
remnants in the experimental data does not affect the results.  We have used
both multiplicity and excitation energy as the control parameter in the
determination of critical exponents and related quantities and find no
difference.  

The nature of the phase transition predicted by SMM has been examined
using the microcanonical equation of state. For
lighter remnants we find evidence of backbending in the caloric curve
and a negative system specific heat, which are the signatures
of a first-order phase transition.  We estimate the transition energy
by means of a Maxwell construction.  The transition energy decreases
with increasing remnant mass and charge and  may extrapolate to zero just above $A
=160$, which might suggest a continuous phase transition in
the breakup of this remnant.  Here the positive system specific heat
peaks at the same $E_i$ value for which both data and SMM
fragment mass yield distributions obey a power law.  
For both the first order and continuous phase transition cases,
SMM$_{hot}$
indicates that the multifragmentation  final state consists of droplets
with $A > 4$ and that particles having $A \leq 4$ account for only
$\sim 8\%$ of the mass in the MF region.

The SMM results agree with the theoretical expectations for a small
($\sim$ 150  constituents) isolated  system.  \cite{gross90,gross97a,gross97}.
The most probable equilibrium state of a highly excited small isolated system with
short range interactions is  an inhomogeneous state. 
 The addition of the long range
Coulomb force lowers the MF transition temperature significantly for
heavier remnants and can influence the order of the thermal phase
transition.  Both the microcanonical  equation of state and thermally
scaled kinetic energy fluctuation arguments favor a continuous phase
transition.

\acknowledgements

We thank Dr. A. S. Botvina for giving us the SMM code and for valuable
discussions concerning its use. This work was supported by the U.S.
Department of Energy under contract numbers (DE-FG02-88 ER 40412C).

\begin{table}
\caption{Parameters in the SMM free energy expression.}
\label{tab1}
\begin{tabular}{lll}
\multicolumn{3}{c}{Fixed Parameters}\\
\tableline\\
$W_0$\hspace{.25in}&Volume binding energy of cold nuclear matter\hspace{.25in}
&16 MeV\\
$\beta_0$\hspace{.25in}&Surface tension of the cold nucleus\hspace{.25in}
&18 MeV\\
$T_c$\hspace{.25in}&Neutral matter critical temperature\hspace{.25in}&18
MeV\\
$\gamma_{sym}$\hspace{.25in}&Symmetry energy coefficient\hspace{.25in}
&25 MeV\\
d&Crack width parameter&1.4 fm\\
$\kappa$\hspace{.25in}&Coulomb reduction parameter\hspace{.25in} &2 \\
\tableline\\
\multicolumn{3}{c}{Single adjustable Parameter}\\
\tableline\\
$\epsilon_0$\hspace{.25in}& Inverse density parameter\hspace{.25in}
&16 MeV\\
\end{tabular}
\end{table}

\begin{table}
\caption{Critical exponents from data and SMM$_{cold}$ (experimental
remnants).}
\label{tab2}
\begin{tabular}{cccccc}
Parameter&Data&SMM\\
\tableline\\
$m_c$&22$\pm$1&20$\pm$2\\
$\tau$&2.19$\pm$0.02$^a$&2.17$\pm$0.02\\
$\gamma$&1.4$\pm$0.3$^a$&1.02$\pm$0.23\\
$\sigma$&0.32$\pm$0.05$^b$&0.63$\pm$0.08$^c$\\
&0.54$\pm$0.11$^b$
\end{tabular}

\begin{description}
\item a.
The results differ slightly from those given in Ref. [10]
because a larger data set was used in the present analysis [58].

\item b.
The first value is obtained by including the largest fragment on the
``gas'' side of $m_c$.  The second value  was
obtained by adjusting  $\sigma$ for the effect of cooling. 
The experimental value for the  scaling
function in Fig. 18a was obtained using the second value of $\sigma$.

\item  c.
This result is obtained by including the largest fragment on the
``gas'' side of $m_c$.
\end{description}
\end{table}

\begin{table}
\caption{Critical exponents from SMM$_{hot}$ and SMM$_{cold}$.}
\label{tab3}
\begin{tabular}{ccccc}
Parameter&
\multicolumn{2}{c}{Experimental remnant}&
\multicolumn{2}{c}{Single remnant} \\
&\multicolumn{2}{c}{distribution} & \multicolumn{2}{c}{A = 160, Z = 64} \\
&SMM$_{cold}$ & SMM$_{hot}$ & SMM$_{cold}$ & SMM$_{hot}$\\
\tableline\\
$m_c$&12$\pm$2&5$\pm$2&12$\pm$2&5$\pm$2 \\
$\tau$&2.17$\pm$0.02&2.05$\pm$0.02&2.07$\pm$0.01&2.03$\pm$0.01 \\
$\gamma$&1.02$\pm$0.23&0.91$\pm$0.20&1.10$\pm$0.20&0.97$\pm$0 .15 \\
$\sigma$&0.62$\pm$0.08&1.04$\pm$0.11&0.69$\pm$0.02&1.00$\pm$0.07 \\
\end{tabular}
\end{table}
\end{document}